# FISH Omics Format: a standard enabling FAIR chromatin tracing data sharing across repositories

***Authors***

Rahi Navelkar [a#], Andrea Cosolo [a#], Iago Maceda Porto [b#], Bogdan Bintu [c], Yubao Cheng [d], Vincent Gardeux [e], Silvia Gutnik [f], Taihei Fujimori [g], Antonina Hafner [h], Atishay Jay [i], Bojing Blair Jia [j], Adam Paul Jussila [k], Gerard Llimos [e], Antonios Lioutas [l], Nuno MC Martins [l], William J Moore [m], Mattia Roccatello [b], Yodai Takei [n], Frances Wong [m], Kaifu Yang [o], Huaiying Zhang [p], Quan Zhu [o], Magda Bienko [q], Lacramioara Bintu [r], Long Cai [s], Bart Deplancke [e], Marcelo Nollmann [t], Susan E Mango [f], Bing Ren [o], Peter J Park [a], Ahilya N Sawh [u], Andrew Schroeder [a], Jason R Swedlow [g], Golnaz Vahedi [i], Chao-Ting Wu [k], Sarah Aufmkolk [k, u#], Alistair N Boettiger [h#], Irene Farabella [b#], Siyuan Wang [d#], Caterina Strambio-De-Castillia [v#],

[#], Contributed equally to this work.

***Affiliations***

a - Department of Biomedical Informatics, Harvard Medical School, Boston, MA 02115, USA
b - Integrative Nuclear Architecture Laboratory, Center for Human Technologies, Istituto Italiano di Tecnologia, Genova, Italy.
c - Shu Chien-Gene Lay, Department of Bioengineering, University of California, San Diego, La Jolla, CA, USA
d - Department of Genetics, Yale University, New Haven, CT 06510, USA
e - Laboratory of Systems Biology and Genetics, Institute of Bioengineering, School of Life Sciences, Ecole Polytechnique Fédérale de Lausanne (EPFL), Lausanne, VD, Switzerland
f - Biozentrum, University of Basel, Basel, BS, Switzerland
g - BioHub, San Francisco, CA
h - Department of Developmental Biology, Stanford University, Stanford, CA 94305, USA
i - Department of Bioengineering, University of Pennsylvania
j - Bioinformatics and Systems Biology Graduate Program, University of California San Diego, La Jolla, CA, USA; Medical Scientist Training Program, University of California San Diego, La Jolla, CA
k - Bioinformatics and Systems Biology Graduate Program, University of California San Diego, La Jolla, CA, USAl - Department of Genetics, Harvard Medical School, Boston, MA 02115, USA
l - Department of Genetics, Harvard Medical School, Boston, MA 02115, USA
m - Divisions of Molecular Cell and Developmental Biology and Computational Biology, University of Dundee, Dundee, UK
n - Biohub, New York, NY, USA
o - Center for Epigenomics, Department of Cellular and Molecular Medicine, School of Medicine, University of California, San Diego, La Jolla, CA, USA
p - Carnegie Mellon University, Department of Biological Sciences, Pittsburgh, PA, 15213, USA
q - Human Technopole, Milan, Italy; Department of Microbiology, Tumor and Cell Biology, Karolinska Institutet, SciLifeLab, Stockholm, Sweden
r - Department of Bioengineering, Stanford University, Stanford, CA 94305, USA
s - Division of Biology and Biological Engineering, California Institute of Technology, Pasadena, CA, USA
t - Center of Structural Biology, Univ Montpellier, CNRS, INSERM, Montpellier, France
u - Department of Biochemistry, University of Toronto, Toronto, Ontario, Canada
v - Program in Molecular Medicine, UMass Chan Medical School, Worcester, MA 01605, USA

The rapid expansion of multiplexed Fluorescence In Situ Hybridization (FISH) methods, collectively termed interchangeably multiplexed FISH or FISH-omics, has transformed our ability to map chromatin organization in single cells, often alongside measurements of RNAs and proteins. Chromatin Tracing experiments [1,2] reconstruct the three-dimensional structure of chromatin fibers across successive imaging rounds and fall into two categories (see Figure 2 in Dekker et al.)[1]: Ball-and-stick [3–38] and volumetric[25,39–42] Chromatin Tracing, which differ from each other in how they visualize the genomic Targeted Segment (Supplemental Table I - Glossary). In ball-and-stick methods, each hybridization round produces fluorescent Spots, each representing an entire targeted genomic segment, typically 2–100 kilobases in length. Volumetric methods instead represent each genomic segment as a cloud of single-molecule localizations, each marking the position of one of the many fluorophores bound to that segment. While the impact of Chromatin Tracing on our understanding of how chromosome architecture relates to transcriptional activity and cell state in health and disease cannot be overstated, processed results from Chromatin Tracing experiments remain difficult to share, reuse, and compare across studies, due to the absence of a standardized data exchange specification—a gap that the broader genomics community addressed decades ago through formats such as BED and VCF.

Here we introduce the FISH Omics Format for Chromatin Tracing (FOF-CT; Figure 1), a community-developed, open file format standard for processed datasets from diverse Chromatin Tracing imaging modalities. FOF-CT is a file format specification, analogous in purpose to FASTA in sequencing studies, and to mmCIF (for atomic coordinates) and CCP4/MRC (for electron density maps) in structural biology, providing rich metadata and consistent annotation. It is not a repository, platform, or analysis tool, but represents a first step toward establishing standards for Chromatin Tracing data. FOF-CT defines how Chromatin Tracing results should be structured, encoded, and annotated, enabling datasets generated across different laboratories, instruments, and protocols to be exchanged, deposited, and reused without manual reformatting. The format is designed to store data efficiently while capturing essential information about data characteristics and provenance.

FOF-CT is organized in a modular fashion around a mandatory core table that defines chromatin traces as ordered lists of individual DNA-FISH bright spots, reporting their XYZ coordinates, BED-specified genomic locations, trace assignments, and cellular region-of-interest allocations. Up to eleven additional linked tables capture spot demultiplexing results, trace-level experimental data, multiplexed RNA-FISH measurements, spot quality metrics, physical landmark coordinates, and cell and sub-cellular region boundaries. All tables share a common structure comprising a machine- and human-readable metadata header, followed by a tabular data section with mandatory and optional columns. Full specifications are available on ReadTheDocs [43]. FOF-CT now natively supports both modalities: FOF-bas-CT captures ball-and-stick Spot/Trace data, while a parallel FOF-vol-CT specification captures volumetric data as tables of individual Localization events (Figure 1).

Adoption of FOF-CT is showcased in over 212 public imaging datasets associated with 28 publications that have been contributed by both 4DN-affiliated and independent laboratories alike, and deposited in FOF-CT format across repositories including the 4D Nucleome (4DN) Data Portal and the third-party Open Microscopy Environment (OME) Image Data Resource (IDR), spanning chromatin tracing experiments in human, mouse, *C. elegans*, and *Drosophila,* and use techniques including multi-omic ORCA, Hi-M, MINA, MERFISH, and seqFISH. Both repositories allow chromatin tracing datasets to be easily identified and have integrated FOF-CT terms into their analysis result tables. In addition, IDR datasets are queryable by chromosome position and image coordinates through the IDR API, and results can be visualized in a multi-dimensional image viewer (see example at https://idr.openmicroscopy.org/iviewer/?shape=7219651), demonstrating that the format achieves genuine cross-repository interoperability, which is the foundational motivation for its development.

The value of a shared file format lies not only in data exchange but in what it enables downstream. The existence of a common structure has directly catalyzed the development of interoperable visualization and analysis tools. Single-Nucleus Analysis Pipeline for multiplexed DNA FISH (SnapFISH) [44] introduced a computational pipeline for detecting chromatin loops across datasets at multiple genomic resolutions, validated across several independently generated FOF-CT-compliant datasets. Dory performs differential analysis of FOF-CT data, and statistically identifies differential patterns between two groups of chromatin traces [45]. ArcFISH models xyz-axis-specific measurement errors in a unified statistical framework to detect multi-level 3D genome features, benchmarked extensively across diverse FOF-CT datasets from mouse and human samples. Imputation tools, including SnapFISH-IMPUTE[46] and ImputeHiFI[47], leverage the growing pool of standardized public datasets to reconstruct missing loci and integrate orthogonal single-cell modalities. These tools would have been substantially harder to develop—and impossible to benchmark rigorously—without a shared format. A particularly direct demonstration of this catalytic effect is provided by the companion manuscript published alongside this Correspondence, which introduces **Chromatin IMaging Analysis (CIMA**; submitted in parallel to this manuscript**)**, a pioneering modular, open-source Python library for the analysis and, importantly, quality assessment of Chromatin Tracing data.

CIMA natively reads and writes the FOF-CT format, enabling the development and refinement of standardized quality metrics across Chromatin Tracing modalities. This mirrors the model established in sequencing-based genomics and structural biology, where standardized formats (for example, FASTA, mmCIF, or MRC) [48–51] represented the means to allow the coevolution of analytical ecosystems, often produced by orthogonal communities.

The broader lesson is one the genomics community learned early [52]: standardized, open file formats are infrastructure. They lower the barrier to data reuse, enable meaningful cross-study comparisons, support the training and validation of computational and machine learning models, and create the conditions for the development, sharing, and improvement of analytical tools by the community rather than rebuilding them independently in each laboratory. Journals and funders are increasingly requiring FAIR-compliant data deposition; FOF-CT provides the format specification that makes FAIR deposition of Chromatin Tracing data feasible.

We invite laboratories generating Chromatin Tracing data to adopt FOF-CT, and we encourage repositories and tool developers to build on this standard. Specifications, documentation, and example datasets are available on ReadTheDocs [43], 4DN Data Portal, and IDR, respectively.

## Acknowledgments

This project could never have been carried out without the leadership, insightful discussions, support, and friendship of all OME consortium members, with particular reference to Josh Moore, Chris Allan, and Jean Marie Burel. We are massively indebted to the RIKEN community for their fantastic work to bring open science into biology. We would like to particularly acknowledge Shuichi Onami for his friendship and support. We thank all members of the 4D Nucleome Imaging Standard Working Group, Data Coordination and Integration Center, and NIH leadership team for the vision and encouragement that allowed us to get this project started. We are particularly grateful to Burak Alver, Ian Fingerman, and John Satterlee. We thank all members of BioImaging North America (in particular Judith Lacoste, Nikki Bialy, Claire M. Brown, Alex Cotten, Kevin Eliceiri, Alison North, Vanessa Ott), German Bioimaging (in particular Stephanie Wiedkamp-Peters), Euro-Bioimaging (in particular Johanna Bischof, Antje Keppler and Yara Reis) and QUAREP-LiMi (in particular Roland Nitschke, Aastha Mathur and all members of the Working Group 7 - Metadata; quarep.org) for invaluable intellectual input, fruitful discussions, advice and moral support. The authors wish to thank all personnel at their home core facilities and their institutions for unwavering support and material help in the absence of which this work could not have been successfully concluded. We are also indebted to the following individuals for their continued and steadfast support: Alicia Birtz, Roger Davis, Kathy Gemme, Jeremy Luban, and Andrea Sjostedt at the University of Massachusetts Medical School.

This work was supported by funding provided to the following authors:

**WJM, FW, and JRS** - W.J.M., F.W., and J.R.S - OME's work described here was funded by the Wellcome Trust (Ref. 221361/Z/20/Z and 313803/Z/24/Z) and a National Institutes of Health Common Fund 4D Nucleome Program grant UM1HG011593.

**HZ** - H.Z was supported by NIH (U01CA260851)

**QZ** - NIDA/PA-20-272 (Shen)

**MB** - M.B. is supported by the Swedish Research Council (grant. no. 2020-02657), Karolinska Institutet (KI Consolidator Grants 2020), and the European Union (ERC, RADIALIS, GA n. 101088408). Views and opinions expressed are those of the authors only and do not necessarily reflect those of the European Union or the European Research Council Executive Agency. Neither the European Union nor the granting authority can be held responsible for them.

**LB** - L.B. was supported by NIH U01DK127419

**MN** - M.N. was supported by ANR grants ANR-23-CE12-0023-01, ANR-24-EXCI-0002, ANR-24-CE12-2240-02 and by ERC grant 724429.

**SEM** - S.E.M. was supported by SNF 310030_197713 and SNF 320030-227954.

**BR** - B.R. was supported by UM1 HG011585.

**PJP** - P.J.P was supported by U01CA200059.

**ANS** - A.N.S. is supported by the Natural Sciences and Engineering Research Council of Canada (NSERC, RGPIN-2024-04767), a Canada Research Chair in 4D Genomics (CRC, 2023-00193), and a Human Frontier Science Program Research Grant (RGP007/2025).

**GV** - G.V. was supported by the Burroughs Welcome Fund, the Chan Zuckerberg Initiative Award, W. W. Smith Charitable Trust, the Sloan Foundation, and the NIH grants UC4 DK112217, U01 DK112217, R01 HL145754, R01AI168240, U01 DK127768, U01 DA052715.

**CTW** - C.T.W. was supported by NIH/NHGRI (RM1HG011016, UM1HG011593).

**SA** - S.A., was supported by the NIH/NHGRI through RM1HG011016 (Center for Genome Imaging) and UM1HG011593 (4DN).

**ANB** - A.N.B. was supported by NIH (U01DK127419)

**IF** - I.F. was supported by The Giovanni Armenise-Harvard Foundation (Career Development Award, 2022) and by the NIH/NHGRI through RM1HG011016 (Center for Genome Imaging) to CTW.

**SW** - S.W. was partly supported by NIH (UH3CA268202, U01CA260701, R01HG011245, R33CA251037, DP2GM137414, R01CA292936, R01HG012969, R01HG013503), Pershing Square Sohn Cancer Research Alliance, American Federation for Aging Research and Hevolution Foundation.

**CSDC** - C.S.D.C. is supported by NSF grant #2513921 and was supported by grant #2021-244318 (5022) awarded to C.S.D.C. by the Chan Zuckerberg Initiative DAF, an advised fund of Silicon Valley Community Foundation, as part of their Imaging Scientist RFA. Major funding also came from NIH grants #U01CA200059-06, #UM1HG011536-04S1 and #U01CA260699.

## Authors Contributions Statement

Author contributions categories utilized here were devised by the CRediT initiative[53,54].

**Co-first authors** (the following authors contributed equally to this work)
**RN:** Conceptualization, Data curation, Methodology, Project administration, Validation, Writing – review & editing
**AC:** Conceptualization, Data curation, Methodology, Validation, Writing – review & editing
**IM:** Conceptualization, Data curation, Methodology, Software, Validation

**Authors involved in acquiring the CT datasets and curating them to be compatible with FOF-CT** (these authors contributed equally to this work and are listed in alphabetical order by last name)

BB: Data curation, Experiments / Investigation
YC: Data curation, Analysis, Experiments / Investigation, Validation
VG: Data curation, Experiments / Investigation
SG: Data curation, Writing – review & editing
TF: Conceptualization, Data curation, Experiments / Investigation
AH: Data curation, Experiments / Investigation
AJ: Data curation, Experiments / Investigation, Resources
BBJ: Analysis, Experiments / Investigation, Software
APJ: Conceptualization, Data curation, Experiments / Investigation, Resources
GL: Data curation, Experiments / Investigation
AL: Conceptualization, Validation, Visualization
NMM: Conceptualization, Validation, Visualization, Writing – review & editing
WJM: Conceptualization, Data curation, Software, Validation
MR: Analysis, Experiments / Investigation, Validation
YT: Experiments / Investigation, Resources
FW: Conceptualization, Data curation, Validation
KY: Experiments / Investigation
HZ: Experiments / Investigation
QZ: Experiments / Investigation, Resources, Writing – review & editing

**Principal investigators of FOF-CT publications** (these authors contributed equally to this work and are listed in alphabetical order by last name)

MB: Funding acquisition, Experiments / Investigation, Writing – review & editing
LB: Conceptualization, Funding acquisition, Experiments / Investigation, Supervision, Writing – review & editing
LC: Funding acquisition, Experiments / Investigation
BD: Funding acquisition
MN: Data curation, Funding acquisition, Experiments / Investigation, Writing – review & editing
SEM: Funding acquisition, Supervision, Writing – review & editing
BR: Funding acquisition, Supervision
PJP: Funding acquisition, Supervision
ANS: Data curation, Funding acquisition, Software
AS: Data curation, Project administration, Supervision
JRS: Funding acquisition, Project administration, Supervision, Validation, Writing – review & editing
GV: Data curation, Funding acquisition, Experiments / Investigation
CW: Conceptualization, Funding acquisition, Writing – review & editing

**Co-senior authors** (the following authors contributed equally to this work and are listed in alphabetical order by last name)
**SA**: Conceptualization, Methodology, Project administration, Supervision, Validation, Visualization, Writing – original draft, Writing – review & editing
**ANB**: Conceptualization, Data curation, Experiments / Investigation, Methodology, Validation, Writing – original draft, Writing – review & editing
**IF**: Conceptualization, Funding acquisition, Methodology, Project administration, Software, Supervision, Writing – original draft, Writing – review & editing
**SW**: Conceptualization, Data curation, Analysis, Funding acquisition, Experiments / Investigation, Methodology, Project administration, Resources, Supervision, Validation, Writing – original draft, Writing – review & editing

**CS**: Conceptualization, Data curation, Funding acquisition, Methodology, Project administration, Resources, Supervision, Validation, Visualization, Writing – original draft, Writing – review & editing

## Conflict of Interest Statement

The authors listed below have provided the following conflict of interest statements:

**BR** - B.R. is a cofounder of Arima Genomics and Epigenome Technologies
**CW** - The Wu laboratory holds or has patent filings pertaining to imaging and has held a sponsored research agreement with Bruker Inc. CTW is a non-equity-holding co-founder of Acuity Spatial Genomics and, through personal connections to George Church, has equity in companies associated with him.
**SA** - The Wu laboratory holds or has patent filings pertaining to imaging and has held a sponsored research agreement with Bruker Inc. CTW is a co-founder of Acuity Spatial Genomics and, through personal connections to George Church, has equity in companies associated with him, including 10x Genomics and Twist.
**SW** - S.W. is a co-inventor on patent applications related to the chromatin tracing technology.

# Figures

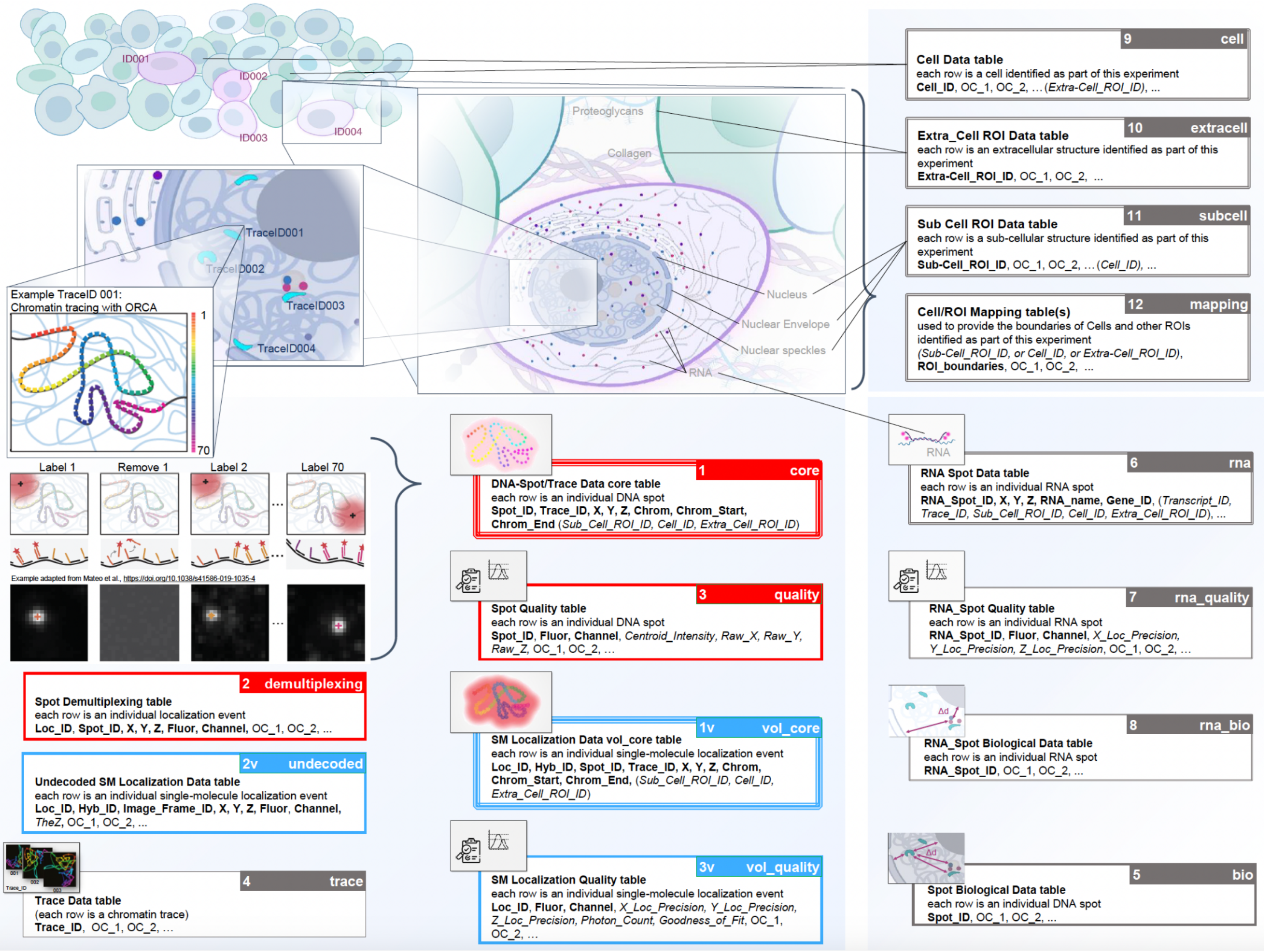

**Figure 1: Schematic Representation of the 15 tables comprising the FISH Omics Format for Chromatin Tracing (FOF-CT).** FOF-CT is a modular format for capturing Chromatin Tracing data from two experimental modalities: Ball-and-Stick and Volumetric. A complete description of the format is available on ReadTheDocs [43]. In this summarized representation, tables are shown as color-coded boxes displaying the table index and abbreviated name in the upper-right corner: *red* for ball-and-stick-specific tables, *blue* for volumetric-specific tables, and *grey* for tables shared across both modalities. Notably, ball-and-stick tables also apply to volumetric Chromatin Tracing, so their use is not restricted to the ball-and-stick modality. In FOF-bas-CT, the mandatory core is the DNA-Spot Data table (Table 1 core), which defines chromatin traces as ensembles of individual DNA-FISH bright Spots (Supplemental Table I - Glossary). Supporting tables provide primary localization data (Table 2 demultiplexing), quality metrics (Table 3 quality), trace information (Table 4 trace), spatial coordinates placing Spots/Traces within the cellular context (Table 5 bio), multiplexed RNA-FISH results (Tables 6 rna, 7 rna_quality, 8 rna_bio), and cellular, extracellular or sub-cellular context (Table 9 cell, 10 extracell, 11 subcell, 12 mapping). When submitting a FOF-bas-CT dataset, deposition of the core Spot/Trace table (Table 1) is mandatory. Demultiplexing (Table 2), quality metrics (Table 3), and biological properties (Table 5) associated with individual Spots are recommended. Trace information (Table 4) is optional, and Tables 6, 7, 8, 9, 10, 11, and 12 are recommended and/or conditionally required depending on the experiment. In FOF-vol-CT, the mandatory core is the Localization Data table (Table 1v vol_core), which associates individual single-molecule localization events (Supplemental Table I - Glossary) with the chromatic coordinates, derived DNA-FISH bright Spots, and traces. Two vol-specific supporting tables capture undecoded single-molecule localization data (Table 2v undecoded) and single-molecule localization quality metrics (Table 3v vol_quality) and are both recommended. All remaining tables available in FOF-bas-CT are also available for FOF-vol-CT and follow the same requirement rules. OC, Optional Column.

# Supplemental Material

The following Supplemental Material is available with this manuscript.

## Abstract

In recent years, multiplexed Fluorescence In Situ Hybridization (FISH) or FISH-omics methods have rapidly expanded, enabling the quantification of chromatin organization in single cells, often in conjunction with measurements of RNAs and proteins. These approaches have deepened our understanding of how 3D chromosome architecture relates to transcriptional activity and cell states in health and disease. Despite these advances, results from Chromatin Tracing FISH-omics experiments remain challenging to share, reuse, and analyze due to the absence of standardized data exchange specifications. Building on the release of microscopy metadata standards, we introduce the FISH Omics Format-Chromatin Tracing (FOF-CT), a community-developed standard for processed results from diverse imaging modalities. We describe the FOF-CT file format and present a curated collection of datasets deposited in the 4D Nucleome Data Portal and the OME Image Data Resource (IDR). We also highlight their potential for reuse, integration, and modeling by outlining example analysis pipelines and illustrating biological insights enabled by standardized, FAIR-compliant Chromatin Tracing datasets. FOF-CT natively represents both ball-and-stick and volumetric Chromatin Tracing data, unifying the two modalities within a single, modular table-based architecture (FOF-bas-CT and FOF-vol-CT, respectively).

## Introduction

Understanding the three-dimensional organization and complex structure of chromatin in four dimensions, incorporating both time and space, is a central goal in modern genome biology, pursued by numerous laboratories and community efforts worldwide, including the 4D Nucleome (4DN) project [1–3]. A significant focus across the field is on mapping the dynamics of chromosome and nuclear organization, as well as the variation observed between single cells and single copies of chromatin. To this end, in addition to sequencing-based techniques such as high-throughput Chromosome Conformation Capture (Hi-C), a variety of microscopy-based Fluorescence In-Situ Hybridization (FISH) techniques are employed.

Classical DNA FISH utilizes fluorescently labelled DNA probes to visually identify specific parts of the genome and capture images of these labeled segments within individual cells, as reviewed by [4–9]. Advances in hybridization protocols [10–17], microfluidics systems [10,13,14,18,19], and the construction of large arrays of barcoded targeting oligonucleotides (including Oligopaints)[16,17,20–23] have expanded the capability of DNA FISH to image multiple DNA segments to potentially cover the entire genome, as reviewed by [5,24–30]. Collectively, these approaches are referred to interchangeably as **Multiplexed FISH** or **FISH-omics**, terminology reviewed by Dekker et al.[2], reflecting their common objective of mapping the positions of numerous (or potentially all) specific DNA segments across the genome, often in conjunction with localizing other cellular components like RNAs and proteins [15,28,31–40]. A variety of DNA FISH Omics protocols have been developed in recent years, where the 3D configuration of chromatin is reconstructed by connecting the location of targeted segments captured in successive imaging rounds, in a genomic mapping process known as "Chromatin Tracing".. Chromatin Tracing techniques can be categorized into two main types (see Figure 2 in Dekker et al.)[2]: **ball-and-stick (bas)** [14,15,31,32,34–65] or **volumetric (vol)** [13,14,66–69]. The distinction between these two approaches lies in how each visualizes the genomic Targeted Segment, and in the nature of the primary imaging observable each produces (Supplemental Table I - Glossary).

In ball-and-stick Chromatin Tracing, each hybridization round labels an entire targeted genomic segment, typically 2–100 kilobases, producing a single, discrete fluorescence signal, or bright Spot. The Spot's spatial position is assigned to the centroid of the fit of the aggregate fluorescence, emitted by the many individual fluorophores bound within the targeted segment. Thus, targeted genomic segments correspond directly to the number of imaged Spots. Ball-and-stick methods, exemplified by the original chromatin tracing technique (Cite Wang 2016) and further developments such as ORCA, Hi-M, MINA, DNA MERFISH, and DNA seqFISH+, have been widely used to date and are well represented among the publicly available Chromatin Tracing datasets described in this manuscript.

By contrast, Volumetric Chromatin Tracing, such as OligoSTORM and OligoDNA-PAINT[13,66,68], is commonly implemented using single-molecule super-resolution imaging techniques to resolve the position of individual fluorophores rather than their aggregate signal. A targeted chromatin segment is therefore visualized as a cloud of many individually resolved Localization events, that collectively samples the volume occupied by the labeled segment and thereby captures its spatial extent. Unlike a ball-and-stick Spot, an individual Localization event does not correspond directly to a single targeted segment, but the cumulative Localization cloud does.

This distinction between a pre-integrated centroid (Spot) and a cloud of individual localization events (Localizations) is not merely descriptive. Rather it reflects a fundamental difference in the primary data each modality generates, and therefore in how each is captured by FOF-CT. To accommodate both without forcing one modality's data model onto the other, FOF-CT is organized around two parallel, interoperable core representations: FOF-bas-CT, built around the mandatory DNA-Spot/Trace core table (Table 1), in which each row reports one bright Spot; and FOF-vol-CT, built around the mandatory Localization Data vol_core table (Table 1v), in which each row reports one Localization event together with its Spot and Trace assignment (see FOF-vol-CT description below). Both core tables share the same downstream architecture, Trace, Cell, ROI, and RNA-FISH tables (Tables 2-12) are common to both modalities, so that ball-and-stick and volumetric datasets can be deposited, queried, and compared side by side within one unified format, rather than two disconnected specifications.

The inherently 3D nature of Chromatin Tracing data enables multi-faceted characterization of chromatin folding conformation through a range of quantitative metrics, thereby revealing novel biological and biophysical insights [70]. At the population level, contact frequency or spatial distance matrices derived from Chromatin Tracing closely recapitulate widely used Hi-C contact maps, capturing both quantitative values and qualitative patterns, including key features such as A/B compartments [71], topologically associating domains (TADs), and promoter–enhancer interactions [14,15,34,38,44–47,50–52,54,72,73]. At the single-cell and single-trace level, spatial distance matrices derived from Chromatin Tracing experiments unveil structural variability across all genomic scales, for example, demonstrating variability in physically segregated A/B compartments [13,31,32,34,39,49,52,54] and highly variable TAD-like structures [14,15,32,54] (also referred to as single-cell domains). Notably, such domain variability has been observed even in the absence of loop extrusion and major epigenetic modifications[74], suggesting that additional mechanisms contribute to chromatin organization beyond canonical extrusion-based models [14,52,57,75]. While plaid-patterned contact enrichment, indicative of separation into A/B compartments, is observed at the population level and appears to be conserved across diverse cell types, the single-cell nature of Chromatin Tracing reveals a more distinct behavior. For example, in mature and slowly dividing cells, individual chromosomes in single cells often partition into physically separated A/B compartments [13,31,32,76]. In contrast, it has been shown that in embryonic stem cells, the compartments exhibit a statistical enrichment of contacts, with individual chromosomes in single cells rarely fully segregating all their A and B regions into two physically distinct compartments [35,36,41]. Chromatin Tracing data also enable the analysis of multi-way contacts [14,40,46,48,52,59,73,77], facilitating the discovery of functional cross-TAD and long-range regulatory interactions mediated by multiple loops [46,52,73,59]. A recent chromatin tracing study discovered a chromatin "curl" structure – a chromatin loop with two long (tens of kb) stem regions aligned in parallel with each other (in contrast to conventional chromatin loops/jets with anti-parallel stem regions) outside of the contexts of DNA recombination/transposition [78]. Chromatin Tracing derived spatial distance matrices also enable quantitative assessment of chromosome compaction by comparing whether distances systematically change across conditions [14,34,37,39,44–46,49,62]. Alternatively, compaction can also be estimated from radii of gyration in contiguous ball-and-stick or volumetric chromatin traces [13,14,31,34,37,39,44,45,49,54,62,63,67,69,79]. Chromatin Tracing provides insights into genome-scale spatial organization at a macro scale by quantifying the radial distribution of each locus within the nuclear space in a genome-wide Chromatin Tracing experiment [35,36,62] and revealing the proximity between genomic loci and nuclear landmarks (such as nuclear lamina, nucleoli, and nuclear speckles) or epigenetic compartments [31,32,35,49]. As a result, FISH-Omics has transformed genome organization studies by enabling single-cell visualization and quantification of chromatin architecture across scales, revealing the importance of structural variability [13–15,31,32,34–39,41–52,66–69].

While this rapid technological progress has driven significant advances, it has also led to a lack of standardization. As individual labs optimized their own pipelines for data acquisition, storage, and analysis, a diversity of bespoke formats and workflows emerged, resulting in a fragmented data-format landscape. The resulting proliferation of incompatible data structures makes cross-study comparisons arduous. However, re-analysis of single-cell Chromatin Tracing datasets has resulted in several novel insights into genome organization and function. For instance, Bohrer and Lason [80] showed that gene co-expression is driven by physical proximity rather than genomic distance. Rajpurkar et al. [38] utilized nascent RNA-labeled Chromatin Tracing data from the Drosophila Bithorax Complex to link chromatin architecture to transcriptional state, revealing that redundant structural layers and rare long-range contacts contribute to transcriptional silencing. Goundaroulis et al. [81] examined single-cell traces from a 2-Mb Targeted Region (Supplemental Table I - Glossary) on human chromosome 21 [14] and demonstrated that interphase chromatin is predominantly unknotted at the megabase scale, with most observed knots attributable to localization errors in densely packed regions, highlighting the need for rigorous quality control. Patel et al. used Chromatin Tracing data from human and mouse cells [14,35,36,82] to develop a method for detecting chromatin domain-like structures and boundaries in single-chromosome pairwise distance matrices, even with substantial missing data. Liefsoens et al. [83] introduced Λ-Plot, a spectral approach that, in conjunction with neural networks, detects nested and complex chromatin loop structures at both the single-cell and population levels using datasets from Bintu et al. [14] and Takei et al. [35]. Several studies developing models for reconstructing the 3D genome have leveraged Chromatin Tracing for both model generation and validation (e.g., Refs. [13,84–90]). Recent work by Remini et al. [91,92] showed that chromatin exhibits two distinct scaling regimes across 5 kb to 2 Mb, both in mouse and in human, corresponding to alternating phases that undergo microphase separation. This observation provides a valuable framework to inform and refine polymer physics models of chromatin organization. Additionally, the increasing availability of public single-cell CT datasets further supports the development [88] and validation [93,94] of accurate, multiscale, data-driven 3D chromatin models.

Such efforts have advanced our understanding of genome organization at the single-cell level. However, their broader impact is often limited by inconsistencies in data formats that hinder data portability and reuse. To support reproducibility and facilitate meaningful cross-study comparisons, the adoption of standardized Microscopy Metadata specifications [95,96], data formats and robust quality assessment frameworks is essential. Addressing this critical bottleneck, an international, multi-laboratory working group, convened initially through the 4D Nucleome (4DN) Consortium and since expanded to include numerous independent groups, developed the FISH-Omics Format for Chromatin Tracing (FOF-CT), a standardized framework designed to unify the representation of DNA FISH data across platforms and studies (Figure 1 and Supplemental Figure 1). The rationale for creating the FOF-CT is to enable the large-scale integration of datasets across different model systems, experimental conditions, and analytical modalities; streamline the development of computational tools, including machine learning applications for spatial genomics; and enhance reproducibility by ensuring compliance with FAIR (Findable, Accessible, Interoperable, Reusable) principles. This manuscript describes the development of FOF-CT, a community standardized data format designed to capture and share results from DNA FISH Omics imaging experiments. This format is immediately compatible with all ball-and-stick Chromatin Tracing techniques described above and is designed to be easy-to-extend to support volumetric methods in the future. Its adoption already extends well beyond 4DN-affiliated laboratories, with independent research groups and public repositories such as the OME Image Data Resource (IDR) integrating the format into their workflows.

## An exchange format for the results of FISH Omics Chromatin Tracing experiments

Translating raw Chromatin Tracing images into a standardized, shareable dataset follows a common sequence of steps — signal detection, assignment to genomic targets, trace and region-of-interest reconstruction, table writing, and format validation — shared by both imaging modalities, even though the specific detection step differs between them. Supplemental Figure 2 summarizes this process end-to-end, from raw multi-round imaging data to a deposited, FOF-CT-compliant dataset, indicating where quality metrics are computed and where format compliance is checked. The two modality-specific implementations of this shared pipeline, FOF-bas-CT and FOF-vol-CT, are each described in detail below.

***FISH Omics Format  for Ball-and-Stick Chromatin Tracing (FOF-bas-CT)***

In **ball-and-stick Chromatin Tracing** experiments, varied trace-reconstruction methods are used to connect the localization of individual DNA-FISH bright Spots (Supplemental Table I - Glossary), thereby reconstructing the three-dimensional (3D) path of the chromatin fiber. While a full description of the proposed FISH Omics Format  for Ball-and-Stick Chromatin Tracing (FOF-bas-CT) specifications can be found on ReadTheDocs [97], a short summary is provided here.

FOF-CT is organized in a modular fashion, ensuring standardization of essential Chromatin Tracing data from multiple techniques, while providing the flexibility to capture additional data that may be required only in specific experimental designs (Figure 1; Supplemental Figures 1 and 3). Specifically, a FOF-bas-CT-compliant datasets must include at least one TXT, CSV, or TSV file, they might contain up to 12 files, each containing one of the prescribed interlinked tables (Figure 1; Supplemental Figures 1 and 3). Specifically, the core of the format consists of the **mandatory DNA-Spot/Trace** core table (i.e., *Table 1*) that defines Chromatin Traces as ordered lists of individual DNA-FISH bright Spots. Specifically, this table reports the XYZ coordinates with respect to the microscopy detection system of individually detected DNA bright Spots, the BED [98] specified genomic location of their target Segment (i.e., Chrom, Chrom_Start, Chrom_End), their Trace assignment as well as their Cellular, Extracellular and Sub-Cellular Region of Interest (ROI) allocation, if appropriate. Additional highly recommended tables, which are linked with the Core table by way of deposition-wide or (ideally) globally unique identifiers (UIDs; Figure 2), support the integration of core Chromatin Tracing results with:

1) The list of individual **Localizations** that are used to identify individual bright Spots with combinatorial barcoding (i.e., Table 2 - Spot Demultiplexing).
2) Supplemental Spot properties such as **Quality metrics** and **Physical coordinates** mapping the position of bright Spots with respect to landmarks defining the architectural organization of the cellular and nuclear space (i.e., Table 3 - Spot Quality and Table 5 - Spot Biological Data)[35,36].
3) Global **Trace**-level experimental results, such as the expression level of nascent RNA transcripts associated with a given Trace or overall localization of the Trace with respect to cellular or nuclear landmarks (i.e., Table 4 - Trace Data).
4) The results of **Multiplexed RNA-FISH** experiments conducted in parallel with the main Chromatin Tracing experiment, consolidated across RNA Spot identification, Quality metrics, and Physical coordinates (i.e., Table 6 - RNA Spot Data[32], Table 7 - RNA_Spot Quality, and Table 8 - RNA_Spot Biological Data).
5) Experimental results that are better captured at the global level of **Cell** (i.e., Table 9 - *Cell Data*; e.g., boundaries and volume), **Extracellular Region of Interest (ROI)** (i.e., *Table 10 - Extra-Cell ROI Data table*; e.g., Tissue), and **Sub-cellular ROI** (i.e., *Table 11 - Sub-Cell ROI Data*; e.g., Nuclear feature or Nucleolus)[15,27,31–36].
6) The boundaries of individual Cells and other Extracellular and Sub-cellular ROIs that were identified as part of the experiment (i.e., *Table 12 - Cell/ROI Mapping*).

In keeping with the modular nature of FOF-CT, all tables are organized in the same general manner and must include two mandatory components: 1) a **header section containing metadata fields,** represented as key-value pairs providing essential information about the experiment and table content and each occupying an individual header line; and 2) a **column section reporting the actual experimental results** in a tabular format. The following rules apply to all table items: 1) Metadata field (i.e., keys) and Column names must use the underscore as a word separator (e.g., *RNA_A_intensity*). 2) Metadata fields and Columns can be required, conditionally required, or optional, where conditionally required items are required only when certain conditions are met (e.g., ##intensity_unit= is required any time an intensity metric is reported). In the **header section**, each line is denoted by the hashtag symbol (i.e., #) and must contain an individual metadata field. Specifically, three different types of header metadata fields are possible:

1) **Human-readable metadata**. These lines must be prefixed with "#" and use the **#key: value** format  (e.g., #lab_name: name of the lab where the experiment was performed).
2) **Mandatory machine-readable metadata.** These lines must be prefixed with "##" and follow the **##key=value** format (e.g., ##FOF-CT_version=v0.1).

3) **User-defined column description.** These lines provide the name and description of optional, user-defined data columns. Descriptions must be understandable and sufficient to ensure the interpretation and reproducibility of the results. These lines must be prefixed with "#^" and follow the **#^key: value** format (e.g., #^optional_column_1: optional column 1 description).

In the **column section**, the first column must always be a relevant UID, such as Spot_ID, Trace_ID, Cell_ID, etc. All other columns must either be mandatory, depending on the table type, or provided at the user's discretion and must be mandatorily defined in the file header as described above. Unless specified, the column and row order is at the user's discretion. If an optional column contains no data (i.e., it is not used), it should be omitted.

To promote interoperability across different genomic modifications, FOF-CT provides instructions for reporting the location of DNA Spots and Traces in case the genome under study contains insertions or deletions [97]. In summary, users are required to provide the following information to relate the location of targeted genomic Segments with respect to the modified genome: 1) a description of the nature and location of the genomic modification; 2) a supplemental Variant Call Format (VCF) [99] file to report the nature and location of the genome modification; 3) the name of the inserted or deleted DNA fragment; and 4) the Start and End coordinates of the target chromosome Segment with respect to the insertion or deletion.

***FISH Omics Format for Volumetric Chromatin Tracing (FOF-vol-CT)***

In **volumetric Chromatin Tracing** experiments, the localization of individual Localizations (Supplemental Table I - Glossary) is reported together with their associated centroid Spots and Traces that reconstruct the three-dimensional (3D) paths of chromatin fibers. The following summary of the proposal supplements the full description provided on ReadTheDocs [97].

A FISH Omics Format for Volumetric Chromatin Tracing (FOF-vol-CT)-compliant dataset must include at least one TXT, CSV, or TSV file and may contain up to 15 files, each reporting one of the prescribed, interlinked FOF-vol-CT-specific tables (Figure 1; Supplemental Figures 1 and 4). The core of the volumetric format consists of the mandatory **Localization** vol_core table (i.e., Table 1v), which provides the XYZ coordinates with respect to the microscopy detection system of each individually detected Localization event, the BED92-specified genomic location of its target Segment (i.e., Chrom, Chrom_Start, Chrom_End), and its Spot and Trace assignment, together with its Cellular, Extracellular, and Sub-Cellular Region of Interest (ROI) allocation, if appropriate.

At minimum, the vol_core table (Table 1v) must report 9 required columns, in order: Loc_ID, Spot_ID, Trace_ID, X, Y, Z, Chrom, Chrom_Start, and Chrom_End; the Sub_Cell_ROI_ID, Cell_ID, and Extra_Cell_ROI_ID columns become conditionally required whenever the corresponding structures were identified as part of the experiment. Two additional tables, linked to the vol_core table by way of deposition-wide unique Localization identifiers (Loc_ID; Figure 1), support the integration of core volumetric Chromatin Tracing results with:

1) The raw, undecoded **primary localization data** underlying a list of **Localization** events (i.e., Table 2v - Undecoded Localization).
2) Per-localization **volumetric Quality metrics** (i.e., Table 3v - Localization Quality) mirroring, at the Localization level, the Spot Quality table described above for FOF-bas-CT.

Critically, all 12 FOF-bas-CT tables described above (Tables 1-12) remain fully available to volumetric datasets and are shared across both modalities rather than duplicated. Taken together, these 3 volumetric-specific tables and the 12 shared FOF-bas-CT tables bring the total number of potential FOF-vol-CT tables to 15 (Figure 1 and Supplemental Figure 1), reflecting a single, unified format for both ball-and-stick and volumetric Chromatin Tracing rather than two disconnected specifications.

## Publication and Sharing of FOF-CT datasets and its adoption by the 4DN project and the OME Public Image Data Resource (IDR)

FOF-CT is a platform-independent file format that has been used for datasets hosted by the 4DN Data Portal [100] but is also supported by the OME Public Image Data Resource (IDR) [101]. Its use across independent repositories demonstrates the portability and broader applicability of the FOF-CT format.

Supplemental Figure 5 and Supplemental Table II summarize 212 public FOF-CT compliant datasets[14,15,31,32,34–36,39,40,44,46,47,49,51–65]. Of these, 203 datasets from 26 publications are available through the 4DN Data Portal, while 9 datasets from 1 publication are hosted on IDR.

The datasets were deposited by laboratories that either belonged to the 4DN consortium (Supplemental Figure 5, *light blue*; 168 datasets; 20 publications) or were external to it (Supplemental Figure 5, *dark blue*; 44 datasets; 8 publications), demonstrating adoption of FOF-CT by the broader research community. Each row in Supplemental Table II represents a group of datasets generated for a single publication. These datasets encompass human, mouse, *C. elegans*, and *Drosophila* cells or tissues imaged using Chromatin Tracing methods including multi-omic ORCA, Hi-M, MINA, MERFISH, and seqFISH. Regardless of the experimental technique, the imaging results are standardized using the FOF-CT format (https://data.4dnucleome.org/resources/data-collections/chromatin-tracing-datasets). The following sections describe the adoption of FOF-CT within the 4DN Data Portal and IDR, respectively.

The 4DN data portal [100] is a multimodal repository primarily designed for genomic datasets generated within and outside the 4DN project. Currently, the portal hosts over 2679 publicly available datasets from at least 108 publications, spanning approximately 47 different sequencing assays and 7 imaging techniques. Identified by unique 4DN identifiers, these datasets can be easily browsed, visualized, and freely downloaded from the 4DN data portal (https://data.4dnucleome.org). Metadata are expertly curated, often in collaboration with the authors, and standardized according to the established 4DN data schema and model (https://data.4dnucleome.org/help/user-guide/data-model). For FOF-CT datasets, processed files and curated metadata underwent the standard 4DN submission and quality-control procedures before being made available through the portal (https://data.4dnucleome.org/microscopy-data-overview). Importantly the metadata includes also a detailed description of the hardware configuration of each microscope that was used for image acquisition, which is provided in compliance with the Light Microscopy (LiMi) - Model of Metadata specifications (formerly 4DN-BINA-OME-QUAREP - NBO-Q)[95]. A list of the Micro-Meta App[96,102] produced Microscope-Hardware JSON files associated with each publication is available in Supplemental Table III. To date, 209 FISH-omics datasets are available via the 4DN Data Portal, substantially expanding the community's access to publicly available data.

IDR is a public, image-focused repository that adds value to imaging data for the scientific community through rich metadata curation [101]. This involves linking image data with experimental metadata, biomolecular annotations, sample preparation protocols, imaging modalities, specimen staining and contrast methods, and analysis results, enabling third-party users to interpret, evaluate, and reproduce imaging results. During curation, errors, omissions, and inconsistencies are identified and corrected, supporting adherence to FAIR principles. Capturing and maintaining accurate metadata is essential to ensuring that data remains reliable and reusable. IDR also uses controlled vocabularies and standardized ontologies to improve the interpretation, comparison, and integration of datasets within and across repositories.

To date, IDR has received two submissions containing multiplexed FISH omics data: Payne et al. [43], and Mota et al.[55]. Together, they highlight the need for consistent annotation to make such data reusable alongside other FISH omics studies. For the Mota et al. dataset[55], IDR collaborated with the FOF-CT working group to incorporate FOF-CT standardized terms into the analysis results tables, which correspond to the FOF-bas-CT *core* table (i.e., Table 1; see example IDR results table at https://idr.openmicroscopy.org/webclient/omero_table/43412838/). These tables are assigned a unique namespace in IDR, allowing all such tables to be retrieved through the IDR API. Tabular data in IDR can be queried by row values, enabling users to identify Spots by chromosome position, image coordinates, and other attributes. Each Spot_ID corresponds to a Shape identifier in IDR, allowing the spot to be viewed directly in a multi-dimensional image viewer (see example at https://idr.openmicroscopy.org/iviewer/?shape=7219651). The IDR team encourages submission of additional datasets annotated with FOF-CT terms to broaden community access to publicly available FISH omics data.

Imaging data submission of this type requires collaboration between data generators, subject matter experts, and data curators and has led to significant advancements in developing data models, metadata schemas, and community-wide accepted data exchange standards for imaging data. Recent packages for the analysis and post-processing of multiplexed DNA-FISH data took a step in this direction by making FOF-CT one of their output and input formats (e.g. pyHiM, traceratops, or the Chromatin IMaging Analysis (CIMA) tool; submitted in parallel to this manuscript)[103–105]. As a result, researchers can leverage quality-controlled, standardized, and AI-ready datasets to effortlessly share and transfer data, perform large-scale, complex downstream analysis, and generate visualization tools in a manner previously unattainable.

## The widespread adoption of FOF-CT facilitates data sharing and empowers downstream visualization and analysis of Chromatin Tracing datasets across repositories

Adopting the FOF-CT establishes a standardized framework that ensures the integration and re-analysis of Chromatin Tracing datasets from diverse protocols, sample types, and data modalities. By enforcing structured metadata and consistent coordinate representation, FOF-CT enhances data reuse and facilitates meaningful comparisons across studies and laboratories, thereby supporting cumulative efforts to build larger and more integrative datasets. Notably, the ongoing deposition of FOF-CT-compliant datasets, with over 200 datasets from 28 publications within a four-year period, reflects growing community engagement.

The development of structural inference and detection methods is a practical example of the value of standardized, shareable Chromatin Tracing data. For instance, Jia et al. [56] re-analyzed multiple ball-and-stick Chromatin Tracing experiments [35,36] to introduce a spatial genome aligner for chromatin trace reconstruction that accounts for copy number variations and aneuploidy. Building on early work in chromatin trace reconstruction [106], this tool opens new avenues for exploring chromatin architecture in genomically complex samples. Similarly, Lee et al. developed SnapFISH (Single-Nucleus Analysis Pipeline for multiplexed DNA FISH) [107], a computational approach for detecting chromatin loops across Chromatin Tracing experiments at different genomic resolutions, considering pairwise chromatin interaction data. Their method was validated across multiple datasets [15,35,36,108], underscoring the power of standardized data formats to facilitate robust cross-study analyses. More recently, Yu et al. introduced ArcFISH (Accurate and Robust Chromatin analysis from FISH data) [109] a novel computational approach that explicitly models the three-dimensional, xyz-axis-specific measurement errors inherent in Chromatin Tracing in a unified statistical framework to detect multi-level 3D genome features. By treating each spatial axis separately, ArcFISH fully exploits the rich information contained in the raw traces. Extensive benchmarking on diverse FOF-CT-compliant Chromatin Tracing datasets—including mouse and human samples and multiple imaging protocols—demonstrates that ArcFISH consistently surpasses existing pipelines in accuracy, interpretability, and biological relevance.

In addition, the availability of standardized, shareable Chromatin Tracing datasets has enabled the development of imputation strategies to address missing data, an inherent limitation in high-resolution Chromatin Tracing experiments. Due to limited detection efficiency, particularly in protocols with multiple hybridization rounds, most ball-and-stick datasets with sub-30 kb genomic resolution report detection rates below 80% [15,32,35,37,51,107]. Initial imputation efforts relied on local smoothing strategies, such as averaging pairwise distances across neighboring loci [35] or across single cells [49], and coordinate interpolation from adjacent spots [51]. While effective in specific contexts, these approaches are limited in their ability to capture global structural features.

Recent advances exploit the growing number of publicly available datasets to improve imputation performance. SnapFISH-IMPUTE [110] identifies similar chromosomal conformations across cells to reconstruct missing loci while preserving local features and global heterogeneity. In parallel, ImputeHiFI [111] integrates Chromatin Tracing with orthogonal single-cell modalities (Hi-C, RNA-seq, RNA FISH) to inform imputation based on multimodal similarity. Together, these tools exemplify how increased data sharing enables more sophisticated analysis pipelines and deeper insights into 3D genome architecture.

The availability of single-cell Chromatin Tracing datasets has opened new avenues for the development of quantitative, data-driven 3D genome modeling, for instance, to generate [88] and validate [94] chromatin folding models at multiple genomic scales. These models help bridge the gap between experimental imaging data and population-level chromatin conformations. As more Chromatin Tracing datasets become available in a harmonized format, they will increasingly support the integration of experimental measurements into predictive and mechanistic 3D genome models.

Efforts to standardize Chromatin Tracing data also enable new tools for interactive visualization. The Nucleome Browser [112], originally developed for IDR data, is being explored for integration with FOF-CT datasets on the 4DN Data Portal. Such tools will enhance interactive exploration of Chromatin Tracing data across studies.

The FOF-CT format has the potential to be a powerful tool for searching, visualizing, analyzing, and re-analyzing Chromatin Tracing data, advancing our ability to investigate the spatial organization of chromosomes and gain a mechanistic understanding of 3D genome function.

## Future Directions

In genomics, data sharing is already well-established, with sequencing data routinely disseminated using well established standardized file formats [113,114] including the Browser Extensible Data (BED) [98], and Variant Call Format (VCF) [99]. As a prerequisite for peer review, most journals enforce strict policies for raw sequencing data—such as RNA-seq—requiring adherence to standards set by community initiatives[115] including the Encyclopedia of DNA Elements (ENCODE)[116], the 4DN project, and others [115,117], and their deposition into public repositories like the Gene Expression Omnibus (GEO) or Sequence Read Archive (SRA). This ensures data transparency, reproducibility, and long-term accessibility, and have since become standard expectations in genomics, proteomics and structural biology research [118,119]. The value of these standardized, open-access resources is well-documented: they foster collaboration, accelerate discovery, and drive the development of novel analytical and visualization tools. A striking example of their impact is the 2024 Nobel Prize in Chemistry [120], which recognized breakthroughs in computational protein design and structural prediction—advancements that relied heavily on large, high-quality, and standardized datasets for training predictive algorithms.

However, no such uniform requirement exists for genomic imaging data. As a result, imaging data is often summarized in figure panels without providing access to the underlying data. Consequently, FISH-based image data are not yet routinely made publicly available upon publication, despite their critical role in elucidating chromatin architecture and its implications for gene regulation, development, and disease. The field is rapidly evolving, with an expanding repertoire of FISH Omics techniques, including LoopTrace [19,121,122], Genome Oligopaint via Local Denaturation (GOLD) FISH [123], Nanoscopy-compatible Oligonucleotides with dyes in Variable Arrays (NOVA) probes [124], Tn5-FISH [125], FRET-FISH [126], or pooled genetic screening methods such as Perturb-tracing [63]. However, this rapid growth also raises concerns about data fragmentation, underscoring the urgent need for shared, extensible, and interoperable exchange formats to integrate diverse datasets into a unified framework.

To address this challenge, a community-driven Imaging Working Group, initially organized under the 4D Nucleome (4DN) initiative, developed FOF-CT, a standardized format designed to facilitate high-quality FISH Omics data production and FAIR data exchange [127]. Its modular architecture (Figure 1, Supplemental Figures 1, 3 and 4) and clear workflow (Supplemental Figure 2) ensures adaptability to both established and emerging multiplexed FISH techniques, making it a versatile solution for the field. Additionally, the modular, table-based design is compatible with the SpatialData format (recently introduced for large-scale spatial omics in biological tissues)[128], thereby paving the way for future integration of spatial omics across scales. The impact of FOF-CT has been profound: since its introduction, adoption has surged, as evidenced by the exponential increase in FOF-CT-compliant datasets deposited in the 4DN Data Portal and on the IDR (Supplemental Figure 5; Supplemental Table II). This widespread, cross-institutional adoption highlights both the unmet need that FOF-CT addresses and its pivotal role in advancing 3D genomics research beyond any single consortium.

Beyond format standardization, establishing shared quality-assessment practices for Chromatin Tracing data remains an important future direction, and one the format is already positioned to support. Consistent with established sequencing formats such as FASTQ and BED, FOF-CT does not enforce a quality threshold at deposition: datasets are currently accepted provided they derive from published or concurrently reviewed work, independent of their quality metrics. Depositors are instead encouraged to report quality metrics so that downstream users can assess and filter each dataset according to the requirements of their own study. The Spot Quality and Localization Quality tables define a concrete set of conditionally required metrics — Goodness_of_Fit, Photon_Count, Peak_Intensity and Centroid_Intensity, X/Y/Z_Loc_Precision and _Loc_Error, and Drift and Chromatic_Shift corrections — whose names and definitions are aligned with the Light Microscopy Model (LiMi-Model) metadata specification, so that Spot- and Localization-level quality reporting remains consistent with broader microscopy metadata standards[95]. Beyond the per-Spot and per-Localization metrics suggested, trace-level quality assessment — for instance, quantifying detection efficiency across successive hybridization rounds — remains an active area of methods development, recently reviewed in [28]. Tools implementing complementary QC approaches are now emerging. For example, CIMA provides polymer-physics-informed assessments of trace consistency based on distances between successive targeted segments, together with measures of detection efficiency that enable quantitative comparisons across datasets. These approaches provide an optional QC layer alongside optical metrics. As methods and community requirements evolve, future versions of FOF-CT may incorporate additional standardized measures, following a model similar to the progressive expansion of the PDBx/mmCIF framework and its associated validation criteria.

The functional study of how nuclear spatial organization influences health and disease demands that we combine sequencing-based methods (e.g. Hi-C) with imaging techniques (e.g. Chromatin Tracing) across many tissue types, cellular states (e.g., cell-cycle phases, differentiation stages, responses to stress), experimental conditions, and labs. Because this data is so varied and complex, interpreting both sequencing- and imaging-based chromatin spatial organization data requires a direct assessment of how "nuclear topography" affects nuclear function. Consequently, efforts initiated by 4DN are underway to create a standardized nuclear reference coordinate system to embed within the FOF-CT. This common coordinate framework (CCF) would anchor imaged chromosome regions within the nucleus, enabling easier data integration, supporting predictive modeling, and increasing the utility of both Hi-C and Chromatin Tracing datasets for computational researchers and the broader scientific community.

The development of FOF-CT represents more than a technical milestone—it signifies a maturation of the DNA FISH omics field. By establishing a common language for Chromatin Tracing data, FOF-CT has eliminated interoperability barriers, fostered collaboration, and accelerated computational innovation. In doing so, it has emerged as the gold standard for multiplexed DNA FISH data, ensuring that the full potential of this transformative technology can be realized.

## Supplemental Figures

***Supplemental Figure 1***

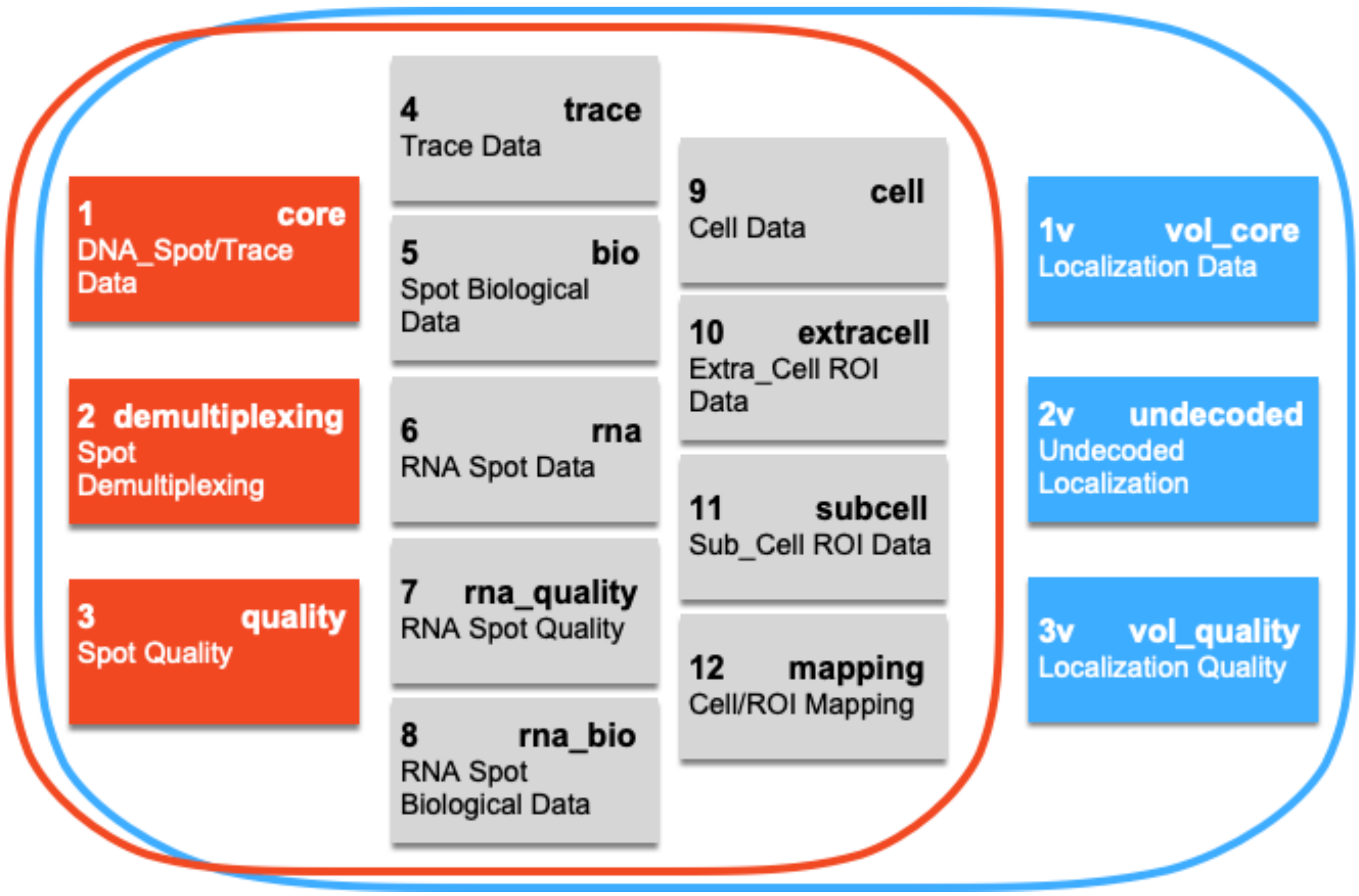

**Supplemental Figure 1. Table architecture of the FISH Omics Format for Chromatin Tracing (FOF-CT), showing which tables are associated with each imaging modality.** The FOF-CT data model comprises 15 tables and is represented here in a nested diagram in which the inner (*red*) and outer (*blue*) circles characterize the ball-and-stick and volumetric Chromatin Tracing modalities, respectively. The three primary tables allowing to capture spot-level ball-and stick Chromatin Tracing data are shown in *red*: DNA-Spot/Trace Data (Table 1-core), Spot Demultiplexing (Table 2-demultiplexing), and Spot Quality (Table 3-quality). The remaining 9 (*gray*) provide supporting Trace-, Cell-, ROI-, and RNA-level information. All 12 *red* and *grey* tables are shared with FOF-vol-CT, which additionally includes three vol-specific tables (*blue)*: Localization Data (Table 1v-vol_core), Undecoded Localization (Table 2v-undecoded), and Localization Quality (Table 3v-vol_quality). Parenthetical labels indicate the names used in the FOF-CT specification.

***Supplemental Figure 2***

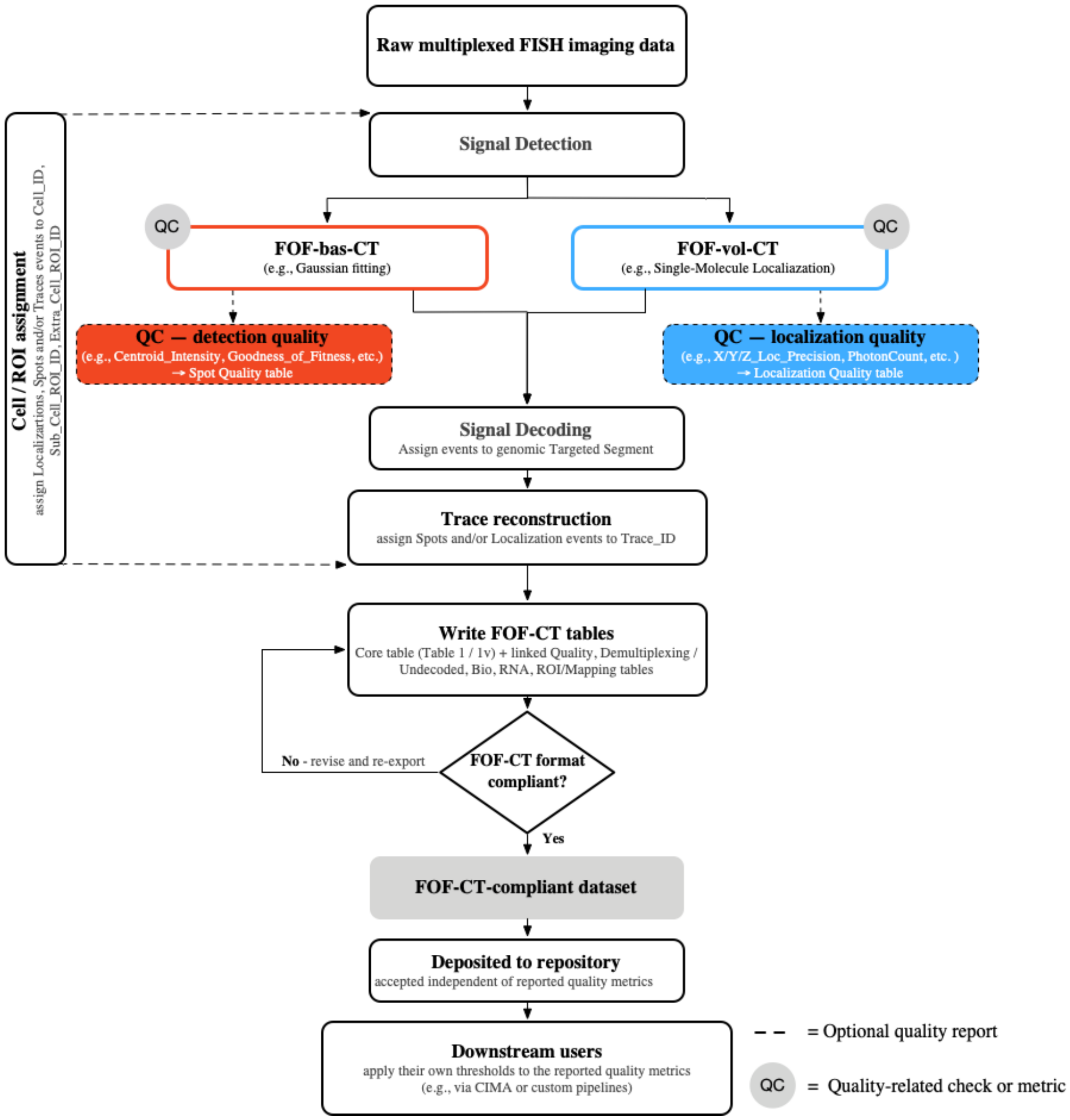

**Supplemental Figure 2. From raw FISH omics images data to a FOF-CT-compliant, deposited dataset.**
Raw FISH omics images undergo signal detection, using methods that depend on the Chromatin Tracing modality: for example, Gaussian fitting for FOF-bas-CT or single-molecule PSF fitting for FOF-vol-CT. Quality metrics computed at this step (e.g., Goodness_of_Fit, Photon_Count, Loc_Precision) are recorded in the corresponding Quality tables but do not gate the pipeline. Detected events are then assigned to targeted genomic segments and assembled into Traces. Depending on the analysis algorithm, allocation to individual Cells and ROI structures can occur at the Localization, Spot, or Trace level. The full set of FOF-CT tables is then written with all required headers and columns. Before deposition, a structural compliance check verifies that the required columns, header fields, and controlled vocabularies are present. This is the only point at which FOF-CT gates a submission. Once compliant, a dataset is deposited in a public repository regardless of its reported quality metrics. Downstream users apply their own quality thresholds to these metrics when deciding which datasets to include in their analyses.

***Supplemental Figure 3***

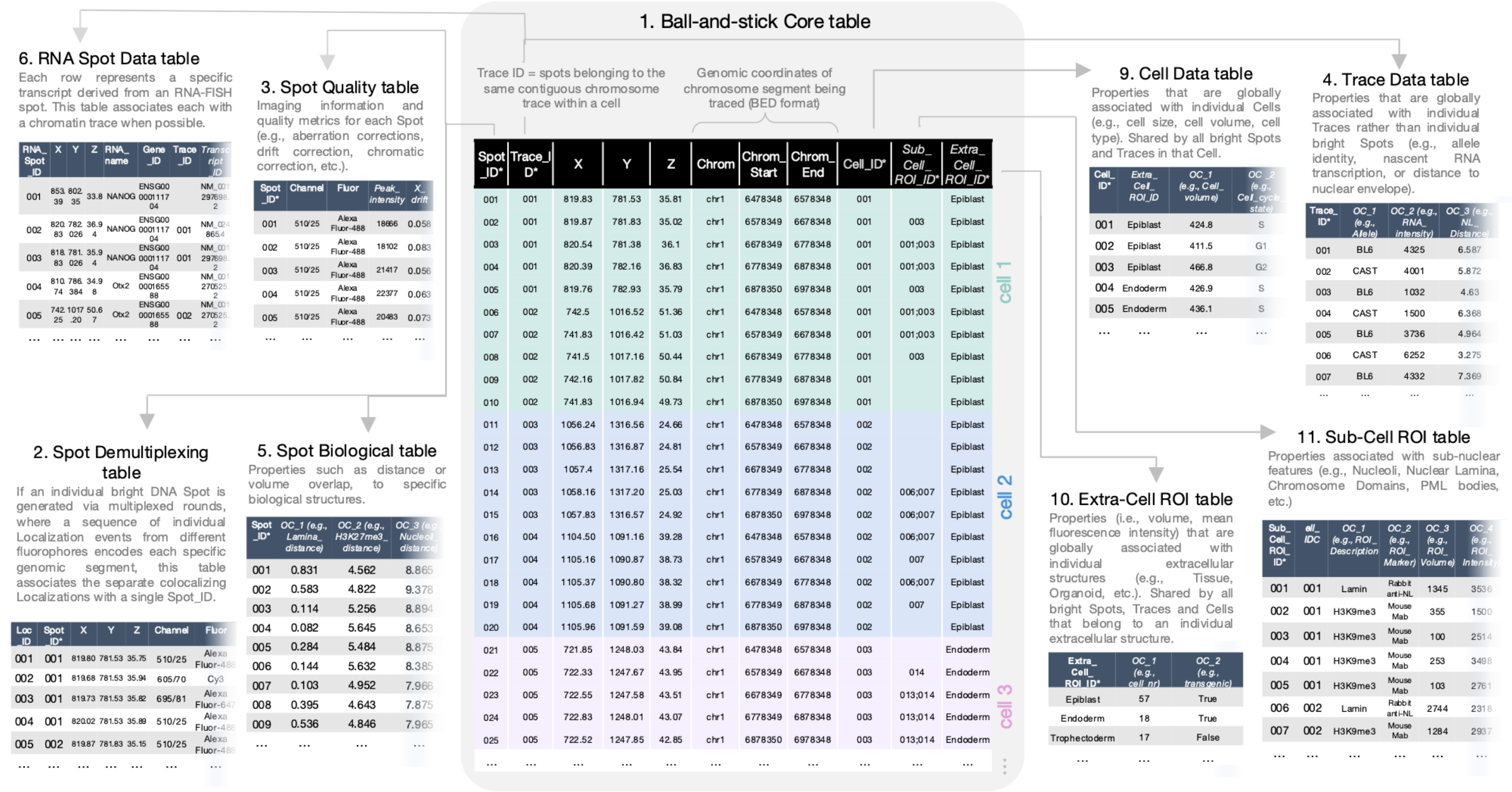

**Supplemental Figure 3. Anatomy of the mandatory core FOF-bas-CT table and connection to other data tables.** The example ball-and-stick (bas) Core Table shown contains the first five traces, corresponding to three cells. Additional example tables are also shown, each linked to the Core Table through a shared identifier column, marked with an asterisk (e.g., Spot_ID links the Spot Quality table to the Core Table, and Cell_ID links the Sub-Cell ROI table to the Core Table). For each example table, only the top few rows are shown; the full table continues beyond what is displayed. Mandatory columns are labeled in bold. Optional Columns (OC) are labeled in italics.

***Supplemental Figure 4***

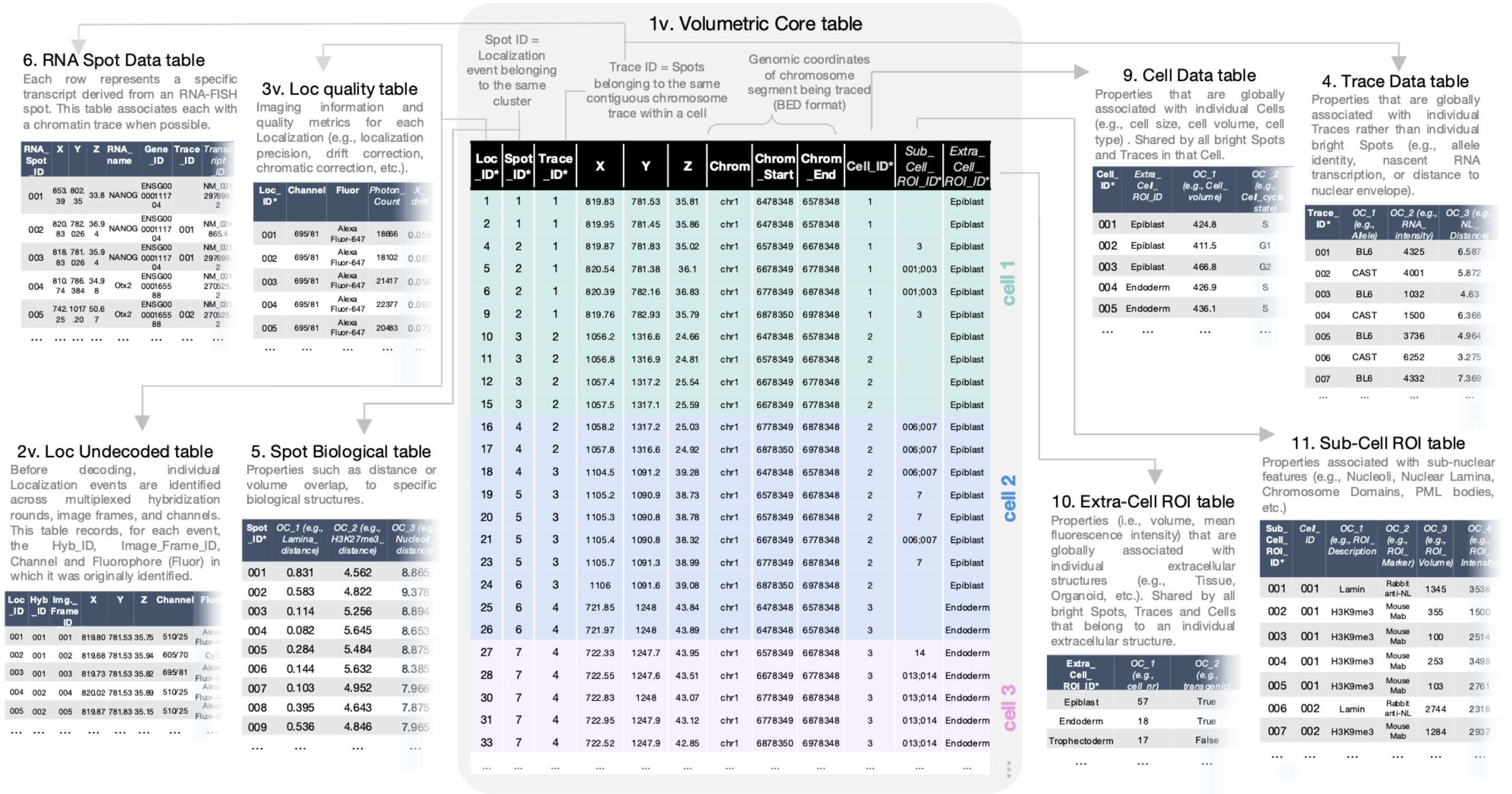

**Supplemental Figure 4. Anatomy of the mandatory core FOF-vol-CT table and connection to other data tables.** The example volumetric Core Table shown contains the first four traces, corresponding to three cells. Additional example tables are also shown, each linked to the Core Table through a shared identifier column, marked with an asterisk [e.g., Loc_ID links the Localization Quality table to the Core Table, and Cell_ID links the Sub-Cell ROI table to the Core Table]. For each example table, only the top few rows are shown; the full table continues beyond what is displayed. Mandatory columns are labeled in bold. Optional Columns (OC) are labeled in italics..

***Supplemental Figure 5***

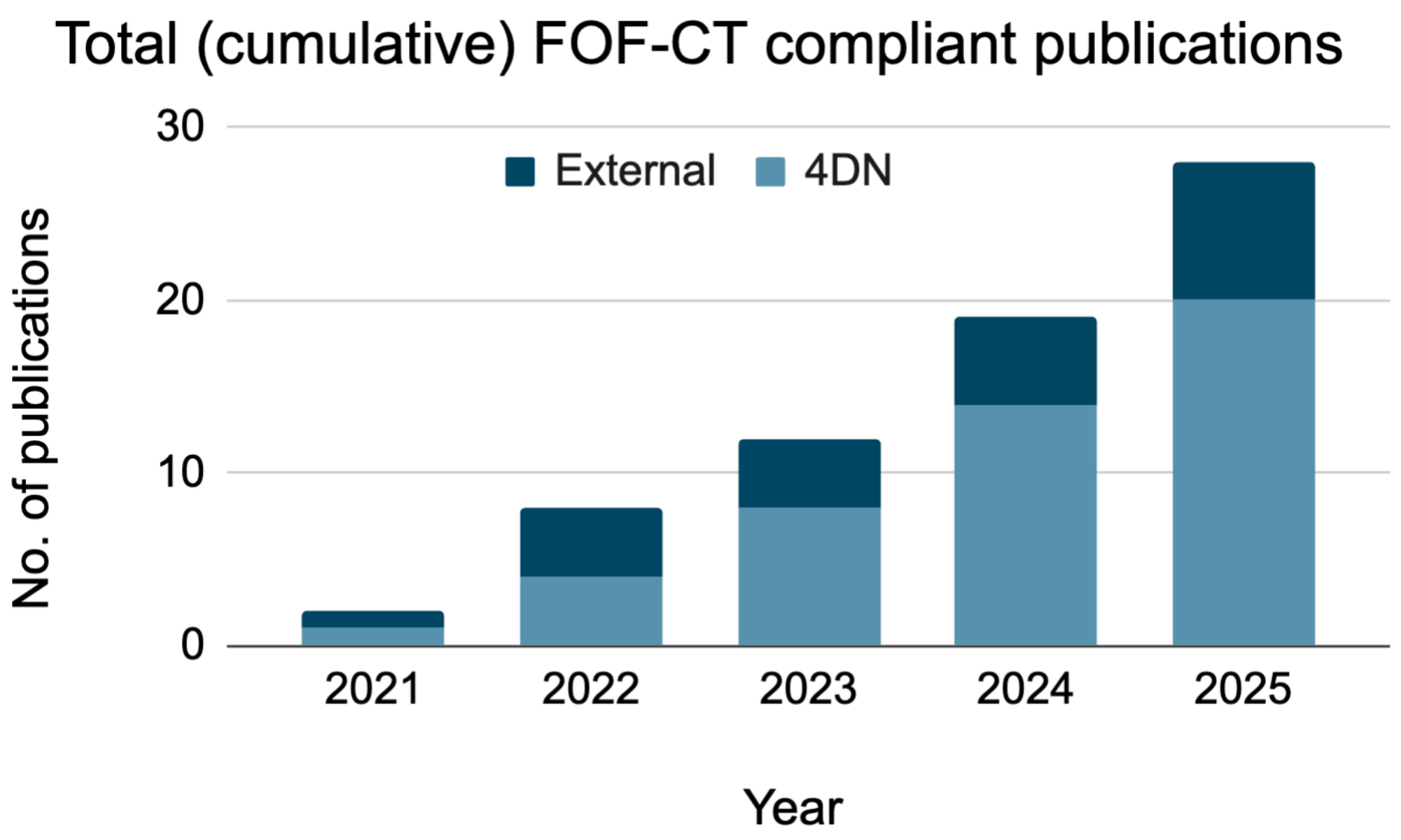

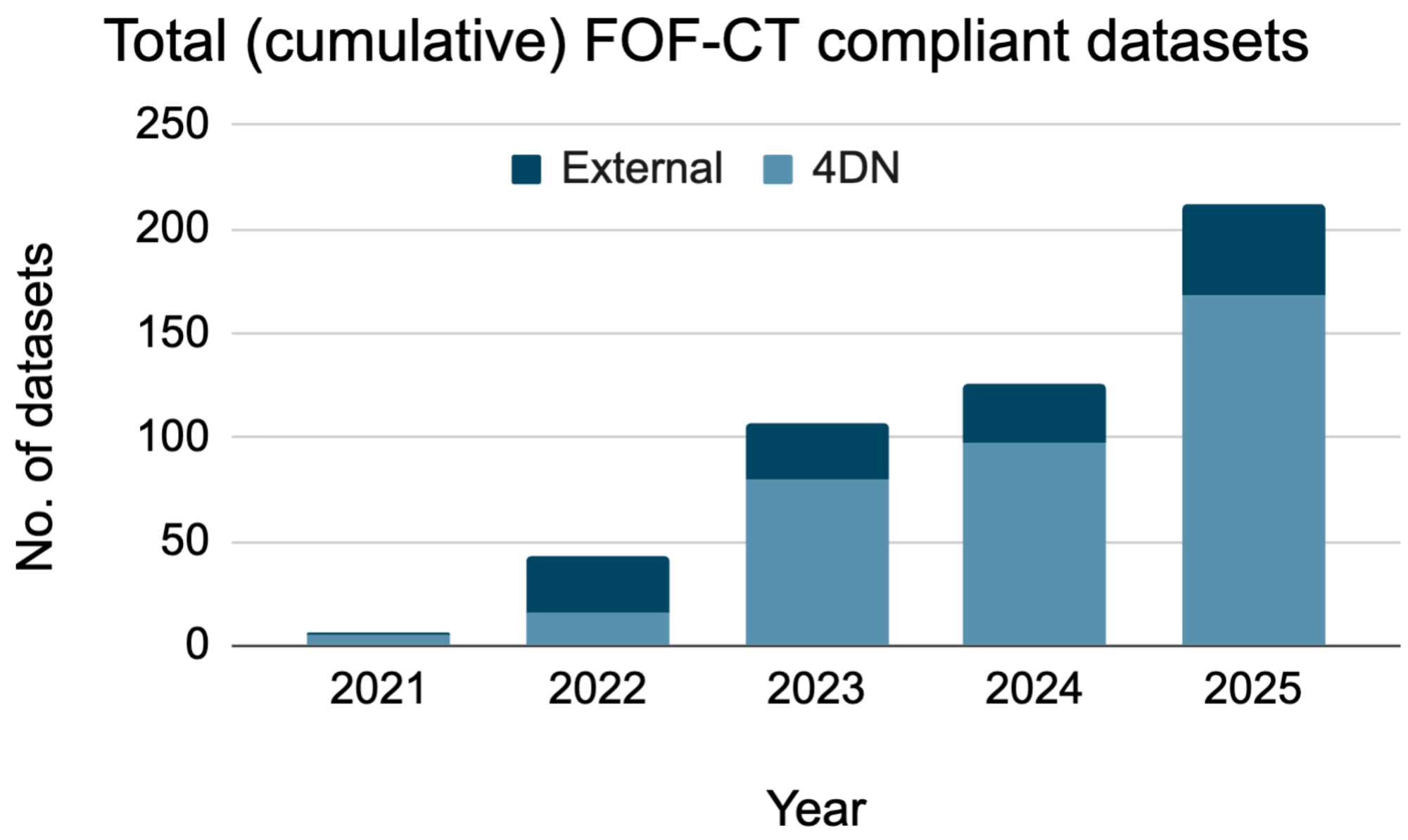

**Supplemental Figure 5. Adoption and impact of the FOF-CT standard on publicly available datasets and publications.** Since its introduction, the FOF-CT standard has driven a substantial increase in compliant datasets and publications (Supplemental Table II) deposited in the 4DN Data Portal and the Image Data Resource (IDR) from both laboratories that belonged to the 4DN consortium (*light blue*) or were External to it (*dark blue*), demonstrating broad community adoption. (*Top*) Number of FOF-CT–compliant publications available through the 4DN Data Portal (n = 26) and the IDR (n = 1; project link: [IDR project 2251]) for a total of 27 publications. (*Bottom*) Number of deposited FOF-CT–compliant datasets in the 4DN Data Portal (n = 203) and the IDR (n = 9; project link: [IDR project 2251]) for a total of 212 datasets. *Note:* Jia et al., 2022 did not generate new datasets but re-analyzed data from Takei et al., 2021 (*Nature*) and Takei et al., 2021 (*Science*).

## Supplemental Tables

***Supplemental Table I - Glossary of Terms***

List of terms used in the main text and their description. Abbreviations: BED, Browser Extensible Data; CT, Chromatin Tracing; BS, Backstreet; MS, Mainstreet.

Supplemental Table I - Glossary.xlsx

Notes: based on Table in slide 2 here and the Glossary of Terms in Slide 9 and 10

***Supplemental Table II: List of FOF-CT compatible datasets and publications that are publicly available on the 4DN Data Portal and IDR***

Detailed summary of FOF-CT compatible publications and associated datasets that are made available to the public as part of this publication.

2025-8-15_Table II-III_ FOF-CT_Dataset overview_RN_CS.xlsx

***Supplemental Table III: Community-driven microscopy Metadata specifications compliant Hardware description of the microscopes used to produce published FOF-CT publications and datasets***

Location on the 4DN Data Portal of the Microscope-Hardware JSON file(s) describing the 4DN-BINA-OME-QUAREP (NBO-Q) Microscopy Metadata specifications[95,96] compliant configurations of the microscopes used to produce the indicated datasets. These Microscope JSON files were produced using the Micro-Meta App software and can be viewed and compared on the 4DN Data Portal using the Micro-Meta App user interface.

2025-8-15_Table II-III_ FOF-CT_Dataset overview_RN_CS.xlsx

## Supplemental Table I: Glossary of Terms

| Name | Description | Alternative names |
|---|---|---|
| **Targeted Region** | A genomic region that is targeted for imaging-based analysis | Locus, DNA region |
| **Targeted Segment** | A segment of genomic DNA or RNA transcript into which the Targeted Region is subdivided to enable imaging-based analysis. The size of the Targeted Segment often determines the genomic resolution of the analysis. | DNA region, RNA sequence, chromosomal segment |
| **Oligo Target** | This is the target sequence for the Target-Specific Oligo. The list of Oligo Target sequences are typically stored in a BED file and are used to costruct the oligo library. Thus if the library has 1000 unique Target-Specific Oligos, there would be 1000 unique OligoTarget sequences. | |
| **Targeting Oligo Sequence** | This is the section of the Target-Specific Oligo sequence that anneals to the Oligo Target. | Target Annealing Sequence; Region of homology to the DNA or RNA target, genome homology region. |
| **Target-Specific Oligo** | This oligonucleotide binds to a unique Oligo Target sequence and is typically made up of three parts: (1) Targeting Sequence – located in the center, complementary to the target. (2) Mainstreet Oligo Sequences (MS) – one or more sequences positioned upstream (5′ side) of the Targeting Oligo Sequence. These are usually named, from 5′ to 3′: Universal MS (UMS), MS1, MS2. (3) Backstreet Oligo Sequences (BS) – one or more sequences positioned downstream (3′ side) of the Oligo Targeting Sequence. These are usually named, from 3′ to 5′: Universal BS (UBS), BS1, BS2. | Oligopaint oligo |
| **Multiplexed-FISH Probe** | This is the pool of multiple (i.e., dozens to thousands) individual Target-Specific Oligos, which when combined in a Chromatin Tracing (CT) experiment allow a Targeted Segment to be identified univocally and detected. | |
| **Single-molecule localization** | The sub-pixel localization of each detected emission event of a single-molecule fluorophore. | |
| **Spot** | The centroid fitting of the imaged Spot of a detected fluorescence signal. In ball-and-stick CT imaged Spots reflect the totality of detected fluorescent signal from multiple individual fluorescennce-emitting molecules (i.e., fluorophores). Thus in ball-and-stick CT "there is an actual correspondence between the number of targeted genomic segments and the number of imaged spots." | Imaged spot, bright spot, DNA/RNA spot |
| **Localization cloud** | A cluster of single-molecule localizations, each representing the fluorescence signal emitted by an individual fluorophore.<br>For example, in volumetric CT, single-molecule localizations from fluorophores bound to target-specific oligos directed against the same targeted segment are grouped into a single localization cloud. Such clouds can contain anywhere from tens to thousands of individual localizations, enabling computation and analysis of the occupied volumes and overlap fractions at each step. | Localization cluster |

**Legend:** List of terms used in the main text and their description. Abbreviations: BED, Browser Extensible Data; CT, Chromatin Tracing; BS, Backstreet; MS, Mainstreet.

**Table II: List of FOF-CT compatible datasets and publications that are publicly available on the 4DN Data Portal and IDR**

| | | | | | | | | | | | | Coverage | | | | | | | | |
|---|---|---|---|---|---|---|---|---|---|---|---|---|---|---|---|---|---|---|---|---|
| Nr. | Reference | DOI | # of FOF-CT Datasets | FOF-CT Experiment Sets URL | FISH Omics Data Type | FISH Omics Technique | Biological sample (organism) | Target cell/tissue (specimen) | Target chromosome | Main Target molecule (RNA, DNA, | Notes | Size (kb) | Number | Name | Segement Size (kb) | Segment Number | Probe Density (probe/kb) | Distance between Segments (kb) | Additional targets | Cellular / Nuclear Context |
| 1 | Wang, S. et al. Spatial organization of chromatin domains and compartments in single chromosomes. Science 353, 598–602 (2016). | https://doi.org/10.1126/science.aaf8084 | 5 | https://data.4dnucleome.org/publications/6162d287-5782-4f40-aacd-d5da75f0770e/#expsets-table | ball-and-stick-CT | ball-and-stick chromatin tracing | H. sapiens | IMR90 | Chr20. Chr21, Chr22, and ChrX | DNA | | 48000-155000 | 4 | Chr 20; Chr 21; | 100; 100; | 30; 34; | 10 | 1000-4000 | mH2A1 protein | Chromosome territories from |
| 2 | Bintu, B. et al. Super-resolution chromatin tracing reveals domains and cooperative interactions in single cells. Science 362, (2018). | 10.1126/science.aau1783 | 10 | https://data.4dnucleome.org/publications/e6451832-8312-4e74-bd74-742801f726e1/#expsets-table | ball-and-stick CT | ball-and-stick chromatin tracing | H. sapiens | IMR90, K562, A549, HCT116 | Chr21 | DNA | | 1200000-2500000 | 4 | Chr 21 | 30 | n/a | 10 | contiguous | | DAPI |
| 3 | Mateo, L. J. et al. Visualizing DNA folding and RNA in embryos at single-cell resolution. Nature 568, 49–54 (2019). | 10.1038/s41586-019-1035-4 | 7 | https://data.4dnucleome.org/publications/6c007db6-3eba-41d8-9d50-cb7e7c05841f/#expsets-table | ball-and-stick CT | Optical Reconstruction of Chromatin | D. melanogaster; M. musculus | embryo; mESC | dm3[bithorax complex (BX-C); | DNA | | 100–700 | 4 | n/a | 2–10 | 34-110 per region | 2-10 | contiguous | RNA | n/a |
| 4 | Liu, M. et al. Multiplexed imaging of nucleome architectures in single cells of mammalian tissue. Nat. Commun. 11, 2907 (2020). | 10.1038/s41467-020-16732-5 | 1 | https://data.4dnucleome.org/publications/fc9e5d2a-39c9-4c37-8fba-e077532bca74/#expsets-table | ball-and-stick CT | Multiplexed Imaging of Nucleome | M. musculus | fetal liver tissue sections | mmChr19 | DNA | | 100-61000 | 2 | 50 TADs along the entire Chr 19; | 5-100 | 69 | 10-30 | 200-2000; contiguous | RNA | Cell boundary (WGA); Nucleus |
| 5 | Sawh, A. N. et al. Lamina-dependent stretching and unconventional chromosome compartments in early C. elegans embryos. Mol. Cell 78, 96-111.e6 (2020). | 10.1016/j.molcel.2020.02.006 | 7 | https://data.4dnucleome.org/publications/996fc79a-65e1-4527-9e7b-0c4b8b9d7511/#expsets-table | ball-and-stick CT | ball-and-stick chromatin tracing | C. elegans | embryo 2-40-cells | ceChr5, ceChr1 | DNA | | 1300-20000 | 2 | Whole Chr 5 and Chr 1; 22 TADs along Chr 5; 14 TADs | 100 | 13-40 | 10 | ~1000 or contiguous | | Nucleus (DAPI; SYTOX Green) |
| 6 | Su, J.-H., Zheng, P., Kinrot, S. S., Bintu, B. & Zhuang, X. Genome-Scale Imaging of the 3D Organization and Transcriptional Activity of Chromatin. Cell 182, 1641-1659.e26 (2020). | https://doi.org/10.1016/j.cell.2020.07.032 | 3 | https://data.4dnucleome.org/publications/bd20862a-7c4c-4a95-9721-e880181da2d4/#expsets-table | ball-and-stick CT | High-resolution; DNA-MERFISH | H. sapiens | IMR90 | High-resolution[Chr2 | DNA, RNA, Proteins | | variable (whole chromosomes) | 2 | High-resolution[Chr 21, Chr 2]; DNA MERFISH[all human | 50; 100 | High-resolution[650, | 5-7 | High-resolution[contigu | RNA | Hi-Resolution: DAPI;DNA- |
| 7 | Cheng, Y., Liu, M., Hu, M. & Wang, S. TAD-like single-cell domain structures exist on both active and inactive X chromosomes and persist under epigenetic perturbations. Genome Biol. 22, 309 (2021). | 10.1186/s13059-021-02523-8 | 9 | https://data.4dnucleome.org/publications/9c1ed2e4-0c1d-48c1-85fb-a1b5d5a6db8b/#expsets-table | ball-and-stick CT | ball-and-stick chromatin tracing | H. sapiens | IMR90, HTERT-RPE1 | Active and Inactive X Chromosomes | DNA | | 840 | 3 | 3 regions on human ChrX | 30 | 28 | 13.3 | contiguous | mH2A1 protein; XIST, TMSB4X and RPL36A RNA | Xi [macroH2A.1 (mH2A1), and XIST long |
| 8 | Huang, H. et al. CTCF mediates dosage- and sequence-context-dependent transcriptional insulation by forming local chromatin domains. Nat. Genet. 53, 1064–1074 (2021). | 10.1038/s41588-021-00863-6 | 5 | https://data.4dnucleome.org/publications/4f766397-5ead-416e-b891-1c95391dcd77/#expsets-table | ball-and-stick CT | multiplexed FISH | M. musculus | mouse Embrionic Stem Cell | mmChr3 | DNA | | 212.5kb | 1 | n/a | 5kb (+7.5kb insert) | 42 | 26.5 | contiguous | Sox2 RNA | |
| 9 | Takei, Y. et al. Integrated spatial genomics reveals global architecture of single nuclei. Nature 590, 344–350 (2021). | 10.1038/s41586-020-03126-2 | 1 | https://data.4dnucleome.org/publications/745502f5-4f9b-41f5-afb4-a8e95d317e58/#expsets-table | ball-and-stick CT | DNA Sequential Fluorescence In Situ | M. musculus | mESC | mmChr1-19, ChrX | RNA, DNA, Proteins | | variable | 40 | 20 chromosomes with every ~1 Mb coverage; ~1.5 Mb region | 25 | 2460 at 1 Mb resolution for | 150-200 probes/25 kb | ~1000 for 1 Mb resolution; | 70 RNA species (mRNA, ncRNA, | DAPI, H3K9me3, H4K20me3, |
| 10 | Takei, Y. et al. Single-cell nuclear architecture across cell types in the mouse brain. Science 374, 586–594 (2021). | 10.1126/science.abj1966 | 1 | https://data.4dnucleome.org/publications/e01653ae-2783-458a-89c5-be2da572c9f8/#expsets-table | ball-and-stick CT | DNA Sequential Fluorescence In Situ | M. musculus | brain | mmChr1-19, ChrX | RNA, DNA, Proteins | | variable | 40 | 20 chromosomes with every ~1 Mb coverage; ~1.5 Mb region | 25 | 2460 at 1 Mb resolution for | 150-200 probes/25 kb | ~1000 for 1 Mb resolution; | 76 RNA species (mRNA, ncRNA, | DAPI, H3K9me3, H4K20me3, |
| 11 | Llimos, G. et al. A leukemia-protective germline variant mediates chromatin module formation via transcription factor nucleation. Nat. Commun. 13, 2042 (2022). | 10.1038/s41467-022-29625-6 | 2 | https://data.4dnucleome.org/publications/b87c4200-5443-460f-9f1c-2f8092c7cafb/#expsets-table | ball-and-stick CT | ORCA | H. sapiens | homozygous REF (GM12282) | Chr17 | DNA | | 200 | 1 | Axin2 locus. 8 kb step size | 2 nominal; 8 effective (due to | 25 | n/a | contiguous | n/a | n/a |
| 12 | Mota, A., Schweitzer, M., Wernersson, E., Crosetto, N. & Bienko, M. Simultaneous visualization of DNA loci in single cells by combinatorial multi-color iFISH. Sci Data 9, 47 (2022). | 10.1038/s41597-022-01139-2 | 9 | http://idr.openmicroscopy.org/webclient/?show=project-2251 | Single-allele pair-wise distance | miFISH | H. sapiens | RPE | Chr1, Chr2, Chr10 | DNA | # | whole chromosomes or chromosome | 3 chromosomes | | Either ~8kb or ~65kb | 16 probes targetting ~65kb on chr2, 23 | 10-12 | Either every ~3Mb, ~10Mb or ~20Mb | | |
| 13 | Jia, B. B., Jussila, A., Kern, C., Zhu, Q. & Ren, B. A spatial genome aligner for resolving chromatin architectures from multiplexed DNA FISH. Nat. Biotechnol. 41, 1004–1017 (2023). | 10.1038/s41587-022-01568-9 | * not applicable | https://data.4dnucleome.org/publications/04e04160-bbda-4e5b-8d34-3e9c8f460e97/#expsets-table | * not applicable | re-analysis of published FOF-CT datasets | 0 | n/a | n/a | n/a | | n/a | n/a | n/a | n/a | n/a | n/a | n/a | n/a | n/a |
| 14 | Murphy, S. E. & Boettiger, A. N. Polycomb repression of Hox genes involves spatial feedback but not domain compaction or phase transition. Nat. Genet. 56, 493–504 (2024). | 10.1038/s41588-024-01661-6 | 5 | https://data.4dnucleome.org/publications/17cd0dce-562c-49ad-beca-0101c0f8c1dc/#expsets-table | ball-and-stick CT | ORCA | M. musculus | embryonic brain tissue, ESCs, embryonic | mmChr6 | DNA, RNA | | 260 | 1 | Hoxa and Evx locus | 5 | 52 | n/a | n/a | Malat1, Hoxa1 or Hoxa13 and Sox2 mRNA | DAPI |
| 15 | Hafner, A. et al. Loop stacking organizes genome folding from TADs to chromosomes. Mol. Cell 83, 1377-1392.e6 (2023) | 10.1016/j.molcel.2023.04.008 | 33 | https://data.4dnucleome.org/publications/1433f4f6-792e-4ec8-b966-206a0b1d5ff0/#expsets-table | ball-and-stick CT | ORCA | M. musculus | mESCs, mESCs depleted of CTCF, mESCs | mmChr3,mmChr6 | DNA, RNA, Proteins | | 220, 39000, 140000 (mmChr6) and | 6 | Hoxa and Sox2 loci | 30 | 73 (mmChr6) and 93 (mmChr3) | n/a | n/a | multiple mRNA species (ntally regulated plurip-otency genes); | DAPI |
| 16 | Chen, L.-F. et al. Structural elements promote architectural stripe formation and facilitate ultra-long-range gene regulation at a human disease locus. Mol. Cell 83, 1446-1461.e6 (2023). | 10.1016/j.molcel.2023.03.009 | 5 | https://data.4dnucleome.org/publications/8d89b44d-1fae-447f-8115-6a93eda5919d/#expsets-table | ball-and-stick CT | ORCA | H. sapiens | H9 hESCs diff to CNCCs | Chr17 | DNA, protein | | 4000 | 1 | Sox9 regulatory domains and flanking TADs | variable | 80 | n/a | n/a | RAD21, H3K27ac | n/a |
| 17 | Hung, T.-C., Kingsley, D. M. & Boettiger, A. N. Boundary stacking interactions enable cross-TAD enhancer-promoter communication during limb development. Nat. Genet. 56, 306–314 (2024). | 10.1038/s41588-023-01641-2 | 1 | https://data.4dnucleome.org/publications/3a910b1f-5fd7-4684-b2a5-e6f2baaecceb/#expsets-table | ball-and-stick CT | ORCA | M. musculus | embryonic limbs | mmChr13 | DNA | | 750 | 1 | Pitx1, encompassing the downstream PelB enhancer, upstream enhancer Pen, and | 10 | 86 | 25 | n/a | n/a | n/a |
| 18 | Patterson, B. et al. Female naïve human pluripotent stem cells carry X chromosomes with Xa-like and Xi-like folding conformations. Sci. Adv. 9, eadf2245 (2023). | 10.1126/sciadv.adf2245 | 21 | https://data.4dnucleome.org/publications/2fab6d6e-0c82-4ce4-872d-9e69af521936/#expsets-table | ball-and-stick CT | ball-and-stick chromatin tracing | H. sapiens | HPSC | ChrX | DNA | | 155000 | 1 | ChrX | 100 | 40 | 10 | 4000 | XIST RNA, ATRX intron RNA | Xi [XIST lncRNA] |
| 19 | Liu, S. et al. Cell type-specific 3D-genome organization and transcription regulation in the brain. Sci. Adv. 11, eadv2067 (2025). | https://doi.org/10.1126/sciadv.adv2067 | 3 | https://data.4dnucleome.org/publications/de7e2893-e23f-4331-899f-888ee267e3a9/#expsets-table | ball-and-stick CT | DNA and RNA MERFISH | M. musculus | mouse brain (RNA MERFISH | whole genome | DNA, RNA | | variable | 3 | whole genome; marker genes for different MOp | 250000 | 988; 28; | n/a | n/a | mRNA expression to determine cell identities | DAPI |
| 20 | Messina, O. et al. 3D chromatin interactions involving Drosophila insulators are infrequent but preferential and arise before TADs and transcription. Nat. Commun. 14, 6678 (2023). | https://doi.org/10.1038/s41467-023-42485- | 2 | https://data.4dnucleome.org/publications/52f8c166-d604-47c0-8590-b44c787a89fe/#expsets-table | ball-and-stick CT | Hi-M | D. Melanogaster | Nuclear-cycle 14 embryos | dmChr2 | DNA | ^ | n/a | 1 | dpp locus | 3 | 138 | 17 | n/a | n/a | DAPI |
| 21 | Tastemel, M. et al. Context-dependent and gene-specific role of chromatin architecture mediated by histone modifiers and loop-extrusion machinery. bioRxivorg (2025) | https://doi.org/10.1101/2025.02.21.639596 | 11 | https://data.4dnucleome.org/publications/7b338c4e-a510-41ff-b65a-389b1e270c97/#expsets-table | ball-and-stick CT | ball-and-stick chromatin tracing | M. musculus | F123-CASTx129 with Sox2 tags | mmChr3 | DNA, RNA | | ~200 | 1 | Sox2 | 5 | 42 | 26.5 | contiguous | Sox2 mRNA | |
| 22 | Le, D. J., et al. Super-enhancer interactomes from single cells link clustering and transcription. bioRxiv (2024) | https://doi.org/10.1101/2024.05.08.593251 | 2 | https://data.4dnucleome.org/publications/805bb891-8958-4221-941d-d3be90a109e0/#expsets-table | ball-and-stick CT | ball-and-stick chromatin tracing | M. musculus | ES-E14TG2a | multiple | DNA | | n/a | 2 | Super Enhancers (SE) identified as part of Whiley et al., 2013 and | 15 | 376 | n/a | variable | n/a | n/a |
| 23 | Jensen, C. L. et al. Long-range regulation of transcription scales with genomic distance in a gene-specific manner. Mol. Cell 85, 347-361.e7 (2025) | 10.1016/j.molcel.2024.10.021 | 3 | https://data.4dnucleome.org/publications/0a8a0fdd-eafe-44f3-b7a4-744b50a98b16/#expsets-table | ball-and-stick CT | ball-and-stick chromatin tracing | M. musculus | R1 mESCs | mmChr5 | DNA | | 192 | 3 | Prdm8-Fgf5 locus and flanking regulatory domain and TADs | 5 | 38 | n/a | 5 | n/a | n/a |
| 24 | Gutnik, S., et al. Multiplex DNA fluorescence in situ hybridization to analyze maternal vs. paternal C. elegans chromosomes. Genome Biol. 25, 71 (2024). | 10.1186/s13059-024-03199-6 | 6 | https://data.4dnucleome.org/publications/1453b156-6de5-438d-b41a-6ffa53fc57f5/#expsets-table | ball-and-stick CT | ball-and-stick chromatin tracing | C. elegans | embryo 2-40-cells | chr5 | DNA | | 1300-20000 | 2 | Whole Chr 5; 21 TADs along Chr 5; strain-specific whole | 100 | 21 | 10 | ~1000 | | Nucleus (DAPI) |
| 25 | Jay, A. et al. Single-allele chromatin tracing reveals genomic clustering of paralogous transcription factors as a mechanism for developmental robustness in T cells. bioRxivorg (2025) | https://doi.org/10.1101/2025.05.30.656885 | 1 | https://data.4dnucleome.org/publications/ebc8c69e-3e8b-477f-b7e6-0d7e75a188f4/#expsets-table | ball-and-stick CT | ORCA | M. musculus | Double-positive T cells | chr9 | DNA | | 900 | 30 | *Ets1-Fli1* locus on chr9 | 900 | 30 | 150-200 probes / 30 kb | contiguous | N/A | N/A |
| 26 | Liu, M. et al. Tracing the evolution of single-cell cancer 3D genomes: an atlas for cancer gene discovery. bioRxiv (2024) | https://doi.org/10.1101/2023.07.23.550157 | 22 | https://data.4dnucleome.org/publications/e83173f7-95e4-40ad-9748-e13b2a518fc6/#expsets-table | ball-and-stick CT | ball-and-stick chromatin tracing; | M. musculus | LUAD, PDAC | All 20 mouse autosomes | DNA, Proteins | | Variable (whole chromosomes; | 20 autosomes; 30 genes | All mouse autosomes; cis-regulatory regions of 30 genes | 5-60 | 473; 1178 | 7-30 | ~5000; contiguous | GFP, tdTomato, SPC, CD45, CK19 | Nucleus (DAPI); lineage tracing |
| 27 | Fujimori, T. et al. Single-cell chromatin state transitions during epigenetic memory formation. bioRxivorg (2023) | https://www.biorxiv.org/content/10.1101/20 | 22 | https://data.4dnucleome.org/publications/109b2954-d7d8-43a2-b7a2-83e3fddaa5dc/#expsets-table | ball-and-stick CT | ORCA | H. sapiens, M. musculus | 293T, K562, mouse | human Chr 19; human Chr 8; | DNA | | 150; 2000; | 3 | AAVS1; MYC; | 5; 10; | 19; 48; | 50-110; 161-232; | 5; 10; | N/A | DAPI |
| 28 | Cheng, Y. et al. Perturb-tracing enables high-content screening of multiscale 3D genome regulators. bioRxivorg (2023) | https://doi.org/10.1101/2023.01.31.525983 | 15 | https://data.4dnucleome.org/publications/c4d07d85-7bc4-43d4-bc30-ee8b34c1ea8d/#expsets-table | ball-and-stick CT | ball-and-stick chromatin tracing | H. sapiens | A549, H1-hESC, RPE-hTERT | Chromsome 21, 22 | DNA | | 48000-51000 | 2 | Chr21; Chr22 | 100 | 34; 27 | 10 | 1000-2000 | protein | Nucleus (DAPI); cell body (total |

**Legend:**

# To visualize the FOF-CT compatible datasets associated with this publication, navigate to individual Datasets, Click on Attachments and Click on the "Eye" icon next to the DatasetX.bed file. Example: https://idr.openmicroscopy.org/webclient/omero_table/43412838/

* Jia et al., 2022 (dataset #13) did not generate new datasets but re-analyzed data from Takei et al., 2021 (Nature; dataset #9) and Takei et al., 2021 (Science; dataset #10).

^ Alternative data source: https://osf.io/aqtxj/

**Table III: Open Microscopy Model compliant Hardware description of the microscopes used to produce published FOF-CT publications and datasets**

| Nr. | Reference | DOI | Micro-Meta App Microscope 1 | Micro-Meta App Microscope 1 Link | Micro-Meta App Microscope 2 | Micro-Meta App Microscope 2 Link | Micro-Meta App Microscope 3 | Micro-Meta App Microscope 3 Link |
|---|---|---|---|---|---|---|---|---|
| 1 | Wang, S. et al. Spatial organization of chromatin domains and compartments in single chromosomes. Science 353, 598–602 (2016). | https://doi.org/10.1126/science.aaf8084 | Harvard-Zhuang_lab_MINA_STORM1_Olympus_IX-71(Wang-2016) | https://data.4dnucleome.org/microscope-configurations/e9aeab8a-8c8f-4e8d-b673-20c6c53dc447 | | | | |
| 2 | Bintu, B. et al. Super-resolution chromatin tracing reveals domains and cooperative interactions in single cells. Science 362, (2018). | 10.1126/science.aau1783 | Harvard-Zhuang_lab_CT-STORM_Nikon_Eclipse_Ti-U_Bintu-2018_Scope1.json | https://data.4dnucleome.org/microscope-configurations/c8e8a284-c6d0-4b97-9670-822ae7a8296f | Harvard-Zhuang_lab_CT-STORM6_Nikon_Eclipse_Ti-U_Bintu-2018_Scope2 | https://data.4dnucleome.org/microscope-configurations/a2b6ca4e-7de9-484f-b8c3-c868b68f929e | Stanford-Boettiger_lab_ORCA_Nikon_Eclipse_Ti2-U_Scope2 | https://data.4dnucleome.org/microscope-configurations/a2b6ca4e-7de9-484f-b8c3-c868b68f929e |
| 3 | Mateo, L. J. et al. Visualizing DNA folding and RNA in embryos at single-cell resolution. Nature 568, 49–54 (2019). | 10.1038/s41586-019-1035-4 | STANFORD-Boettiger_Lab_ORCA_Nikon_Eclipse_Ti-U_Scope1.json | https://data.4dnucleome.org/microscope-configurations/84d52890-7e46-4cf8-810c-071ed0673ce8 | STANFORD-Boettiger_Lab_ORCA_Nikon_eclipse_Ti2-U_Scope2 | https://data.4dnucleome.org/microscope-configurations/a2b6ca4e-7de9-484f-b8c3-c868b68f929e | | |
| 4 | Liu, M. et al. Multiplexed imaging of nucleome architectures in single cells of mammalian tissue. Nat. Commun. 11, 2907 (2020). | 10.1038/s41467-020-16732-5 | YALE-Wang_Lab_MINA_Nikon_Eclipse_Ti2-U_Jellyfish_Config1.json | https://data.4dnucleome.org/microscope-configurations/d8f51a9c-1e1a-4a6c-9432-ddf7de64e836 | | | | |
| 5 | Sawh, A. N. et al. Lamina-dependent stretching and unconventional chromosome compartments in early C. elegans embryos. Mol. Cell 78, 96-111.e6 (2020). | 10.1016/j.molcel.2020.02.006 | Harvard-Zhuang_lab_MINA_STORM1_Olympus_IX-71(Wang-2016) | https://data.4dnucleome.org/microscope-configurations/e9aeab8a-8c8f-4e8d-b673-20c6c53dc447 | UniBasel_Biozentrum-Mango_lab_CT_Nikon_Ti2_Scope1 | https://data.4dnucleome.org/microscope-configurations/b904ca5d-fcac-4a7d-b703-dc98adedabf5 | | |
| 6 | Su, J.-H., Zheng, P., Kinrot, S. S., Bintu, B. & Zhuang, X. Genome-Scale Imaging of the 3D Organization and Transcriptional Activity of Chromatin. Cell 182, 1641-1659.e26 (2020). | https://doi.org/10.1016/j.cell.2020.07.032 | HMS-Zhuang_lab_CT-STORM_Nikon_Eclipse_Ti-U_Su-2020_Scope1.json | https://data.4dnucleome.org/microscope-configurations/b489483d-e203-45e1-b52f-fe5e8c8538e0 | HMS-Zhuang_lab_CT-STORM_Nikon_Eclipse_Ti-U_Su-2020_Scope2.json | https://data.4dnucleome.org/microscope-configurations/849c69ce-a0b1-4fe7-8337-ebfc7b308622 | | |
| 7 | Cheng, Y., Liu, M., Hu, M. & Wang, S. TAD-like single-cell domain structures exist on both active and inactive X chromosomes and persist under epigenetic perturbations. Genome Biol. 22, 309 (2021). | 10.1186/s13059-021-02523-8 | YALE-Wang_Lab_MINA_Nikon_Eclipse_Ti2-U_Jellyfish_Config1.json | https://data.4dnucleome.org/microscope-configurations/d8f51a9c-1e1a-4a6c-9432-ddf7de64e836 | YALE-WANG_Lab_MINA_Nikon_Eclipse_Ti2-U_Jellyfish_Config2.json | https://data.4dnucleome.org/microscope-configurations/f0172a96-2bf5-4e6a-938d-f205c6e728c9 | | |
| 8 | Huang, H. et al. CTCF mediates dosage- and sequence-context-dependent transcriptional insulation by forming local chromatin domains. Nat. Genet. 53, 1064–1074 (2021). | 10.1038/s41588-021-00863-6 | UCSD-Ren_lab_BASCT_Nikon_Eclipse_Ti2-U_Lemon.json | https://data.4dnucleome.org/microscope-configurations/98ea721e-8fb1-4dac-80b7-aa46b1d0e1e0 | | | | |
| 9 | Takei, Y. et al. Integrated spatial genomics reveals global architecture of single nuclei. Nature 590, 344–350 (2021). | 10.1038/s41586-020-03126-2 | Caltech-Cai_lab_DNASeqFISH+_Leica_DMi8_SpinnigDisk_Takei_2021_Nature- | https://data.4dnucleome.org/microscope-configurations/751988a8-ea7a-4f40-8507-befa2e473f6c | | | | |
| 10 | Takei, Y. et al. Single-cell nuclear architecture across cell types in the mouse brain. Science 374, 586–594 (2021). | 10.1126/science.abj1966 | Caltech-Cai_lab_DNASeqFISH+_Leica_DMi8_SpinnigDisk_Takei_2021_Nature- | https://data.4dnucleome.org/microscope-configurations/751988a8-ea7a-4f40-8507-befa2e473f6c | | | | |
| 11 | Llimos, G. et al. A leukemia-protective germline variant mediates chromatin module formation via transcription factor nucleation. Nat. Commun. 13, 2042 (2022). | 10.1038/s41467-022-29625-6 | STANFORD-Boettiger_Lab_ORCA_Nikon_Eclipse_Ti-U_Scope1.json | https://data.4dnucleome.org/microscope-configurations/84d52890-7e46-4cf8-810c-071ed0673ce8 | | | | |
| 12 | Mota, A., Schweitzer, M., Wernersson, E., Crosetto, N. & Bienko, M. Simultaneous visualization of DNA loci in single cells by combinatorial multi-color iFISH. Sci Data 9, 47 (2022). | 10.1038/s41597-022-01139-2 | KarolinskaInstitutet_Bienko_Lab_CT_Nkon_Eclipse_Ti-E_Scope1(Mota-2022) | https://data.4dnucleome.org/microscope-configurations/71849a75-5a46-41b1-a0f9-59ad4de6628d | | | | |
| 13 | Jia, B. B., Jussila, A., Kern, C., Zhu, Q. & Ren, B. A spatial genome aligner for resolving chromatin architectures from multiplexed DNA FISH. Nat. Biotechnol. 41, 1004–1017 (2023). | 10.1038/s41587-022-01568-9 | not applicable | not applicable | | | | |
| 14 | Murphy, S. E. & Boettiger, A. N. Polycomb repression of Hox genes involves spatial feedback but not domain compaction or phase transition. Nat. Genet. 56, 493–504 (2024). | 10.1038/s41588-024-01661-6 | STANFORD-Boettiger_lab_ORCA_Nikon_Eclipse_Ti2-U_Scope2 | https://data.4dnucleome.org/microscope-configurations/33ea326f-a557-4457-becf-55bc860c4bdf | STANFORD-Boettiger_lab_ORCA_Nikon_Eclipse_Ti2-U_Scope3 | https://data.4dnucleome.org/microscope-configurations/4fc82c6a-a9cd-43e1-aee8-36db3174b409 | | |
| 15 | Hafner, A. et al. Loop stacking organizes genome folding from TADs to chromosomes. Mol. Cell 83, 1377-1392.e6 (2023) | 10.1016/j.molcel.2023.04.008 | STANFORD-Boettiger_lab_ORCA_Nikon_Eclipse_Ti2-U_Scope2 | https://data.4dnucleome.org/microscope-configurations/33ea326f-a557-4457-becf-55bc860c4bdf | STANFORD-Boettiger_lab_ORCA_Nikon_Eclipse_Ti2-U_Scope3 | https://data.4dnucleome.org/microscope-configurations/4fc82c6a-a9cd-43e1-aee8-36db3174b409 | | |
| 16 | Chen, L.-F. et al. Structural elements promote architectural stripe formation and facilitate ultra-long-range gene regulation at a human disease locus. Mol. Cell 83, 1446-1461.e6 (2023). | 10.1016/j.molcel.2023.03.009 | STANFORD-Boettiger_Lab_ORCA_Nikon_Eclipse_Ti-U_Scope1 | https://data.4dnucleome.org/microscope-configurations/84d52890-7e46-4cf8-810c-071ed0673ce8 | | | | |
| 17 | Hung, T.-C., Kingsley, D. M. & Boettiger, A. N. Boundary stacking interactions enable cross-TAD enhancer-promoter communication during limb development. Nat. Genet. 56, 306–314 (2024). | 10.1038/s41588-023-01641-2 | STANFORD-Boettiger_Lab_ORCA_Nikon_Eclipse_Ti-U_Scope1 | https://data.4dnucleome.org/microscope-configurations/84d52890-7e46-4cf8-810c-071ed0673ce8 | STANFORD-Boettiger_lab_ORCA_Nikon_Eclipse_Ti2-U_Scope2 | https://data.4dnucleome.org/microscope-configurations/33ea326f-a557-4457-becf-55bc860c4bdf | | |
| 18 | Patterson, B. et al. Female naïve human pluripotent stem cells carry X chromosomes with Xa-like and Xi-like folding conformations. Sci. Adv. 9, eadf2245 (2023). | 10.1126/sciadv.adf2245 | YALE-Wang_Lab_MINA_Nikon_Eclipse_Ti2-U_Jellyfish_Config1.json | https://data.4dnucleome.org/microscope-configurations/d8f51a9c-1e1a-4a6c-9432-ddf7de64e836 | YALE-WANG_Lab_MINA_Nikon_Eclipse_Ti2-U_Jellyfish_Config2.json | https://data.4dnucleome.org/microscope-configurations/f0172a96-2bf5-4e6a-938d-f205c6e728c9 | | |
| 19 | Liu, S. et al. Cell type-specific 3D-genome organization and transcription regulation in the brain. Sci. Adv. 11, eadv2067 (2025). | https://doi.org/10.1126/sciadv.adv2067 | Harvard-Zhuang_lab_CT-STORM_Nikon_Eclipse_Ti-U_Liu-2023_Scope1 | https://data.4dnucleome.org/microscope-configurations/9a4086c5-ec74-4e84-b00a-7661505e1273/ | | | | |
| 20 | Messina, O. et al. 3D chromatin interactions involving Drosophila insulators are infrequent but preferential and arise before TADs and transcription. Nat. Commun. 14, 6678 (2023). | https://doi.org/10.1038/s41467-023-42485-y | UMontpellier-Nollmann_lab_CT-Hi-M_ASI_RAMM_Scope1_Config2(Messina-2023) | https://data.4dnucleome.org/microscope-configurations/450f8ca9-4dfa-4fc2-9886-8860a79324f0 | | | | |
| 21 | Tastemel, M. et al. Context-dependent and gene-specific role of chromatin architecture mediated by histone modifiers and loop-extrusion machinery. bioRxivorg (2025) | https://doi.org/10.1101/2025.02.21.639596 | UCSD-Ren_lab_BASCT_Nikon_Eclipse_Ti2-U_Lemon.json | https://data.4dnucleome.org/microscope-configurations/98ea721e-8fb1-4dac-80b7-aa46b1d0e1e0 | | | | |
| 22 | Le, D. J., et al. Super-enhancer interactomes from single cells link clustering and transcription. bioRxiv (2024) | https://doi.org/10.1101/2024.05.08.593251 | STANFORD-Boettiger_Lab_ORCA_Nikon_Eclipse_Ti-U_Scope1 | https://data.4dnucleome.org/microscope-configurations/84d52890-7e46-4cf8-810c-071ed0673ce8 | | | | |
| 23 | Jensen, C. L. et al. Long-range regulation of transcription scales with genomic distance in a gene-specific manner. Mol. Cell 85, 347-361.e7 (2025) | 10.1016/j.molcel.2024.10.021 | STANFORD-Boettiger_Lab_ORCA_Nikon_Eclipse_Ti-U_Scope1 | https://data.4dnucleome.org/microscope-configurations/84d52890-7e46-4cf8-810c-071ed0673ce8 | | | | |
| 24 | Gutnik, S., et al. Multiplex DNA fluorescence in situ hybridization to analyze maternal vs. paternal C. elegans chromosomes. Genome Biol. 25, 71 (2024). | 10.1186/s13059-024-03199-6 | UniBasel_Biozentrum-Mango_lab_CT_Nikon_Ti2_Scope1 | https://data.4dnucleome.org/microscope-configurations/b904ca5d-fcac-4a7d-b703-dc98adedabf5 | UniBasel_Biozentrum-Mango_lab_CT_Nikon_Ti2_Scope2 | https://data.4dnucleome.org/microscope-configurations/94d84b2f-1b93-4025-95fa-acaeb46bfea9 | | |
| 25 | Jay, A. et al. Single-allele chromatin tracing reveals genomic clustering of paralogous transcription factors as a mechanism for developmental robustness in T cells. bioRxivorg (2025) | https://doi.org/10.1101/2025.05.30.656885 | UPenn-Vahedi Lab-BPFPALM-Vutara VXL-Scope1 | https://data.4dnucleome.org/microscope-configurations/e1ea4378-4231-4d6e-ad8f-47cc67ebf284 | | | | |
| 26 | Liu, M. et al. Tracing the evolution of single-cell cancer 3D genomes: an atlas for cancer gene discovery. bioRxiv (2024) | https://doi.org/10.1101/2023.07.23.550157 | Yale-Wang_lab_MINA_Nikon_Eclipse_Ti2-U_Nautilus_Config2 | https://data.4dnucleome.org/microscope-configurations/1fe89e41-239b-4f0a-97c2-9303b3a4bb0b/ | Yale-Wang_lab_MINA_Nikon_Eclipse_Ti2-U_Jellyfish_Config2 | https://data.4dnucleome.org/microscope-configurations/f0172a96-2bf5-4e6a-938d-f205c6e728c9/ | | |
| 27 | Fujimori, T. et al. Single-cell chromatin state transitions during epigenetic memory formation. bioRxivorg (2023) | https://www.biorxiv.org/content/10.1101/2023.10.03.560616v1 | Stanford-Bintu_lab_ORCA_Leica_DMi8_Fujimori_2023_BiorXiv_Scope1 | https://data.4dnucleome.org/microscope-configurations/802e9e45-4ac0-4c4c-b991-3f0a8eced063/ | | | | |
| 28 | Cheng, Y. et al. Perturb-tracing enables high-content screening of multiscale 3D genome regulators. bioRxivorg (2023) | https://doi.org/10.1101/2023.01.31.525983 | Yale-Wang_lab_MINA_Nikon_Eclipse_Ti2-U_Hydra | https://data.4dnucleome.org/microscope-configurations/e72da879-82a1-4dab-b9f9-ba5f8dc92e0a/ | Yale-Wang_lab_MINA_Nikon_Eclipse_Ti2-U_Adorabilis | https://data.4dnucleome.org/microscope-configurations/db4ff6b1-62d1-44b0-8954-bdb0f7477006/ | Yale-Wang_lab_MINA_Nikon_Eclipse_Ti2-U_Kraken | https://data.4dnucleome.org/microscope-configurations/dee759ad-4463-4f49-880d-1ba9ef69f084/ |

**Legend:**
Listed here are the location on the 4DN Data Portal of the Microscope-Hardware JSON file(s) describing the Open Microscopy Model (OMM) compliant configurations of the microscopes used to produce the indicated datasets.
These Microscope JSON files were produced using the Micro-Meta App software and can be viewed and compared on the 4DN Data Portal using the Micro-Meta App user interface.